\documentclass[sn-mathphys]{sn-jnl}% Math and Physical Sciences Reference Style
\jyear{2022}%

\theoremstyle{thmstyleone}%
\theoremstyle{thmstyletwo}%

\theoremstyle{thmstylethree}%

\usepackage{xcolor}
\usepackage[utf8]{inputenc}
\begin{document}

\title[Article Title]{Lagging Heat Models in Thermodynamics and Bioheat Transfer: a Critical Review}


\author*[1]{\fnm{Zahra} \sur{Shomali}}\email{shomali@modares.ac.ir}

\author[2,3]{\fnm{R\'obert} \sur{Kov\'acs}}\email{kovacs.robert@wigner.hu}

\author[2,3]{\fnm{P\'eter} \sur{V\'an}}\email{van.peter@wigner.hu}

\author[4]{\fnm{Igor} \sur{Vasilievich Kudinov}}\email{igor-kudinov@bk.ru}

\author[5]{\fnm{Jafar} \sur{Ghazanfarian}}\email{j.ghazanfarian@znu.ac.ir}

\affil*[1]{\orgdiv{Department of Physics, Faculty of Basic Sciences}, \orgname{Tarbiat Modares University}, \orgaddress{\city{Tehran}, \postcode{14115-175}, \country{Iran}}}

\affil[2]{\orgdiv{Department of Theoretical Physics}, \orgname{Institute for Particle and Nuclear Physics, Wigner Research Centre for Physics}, \orgaddress{\city{Budapest}, \country{Hungary}}}

\affil[3]{\orgdiv{Department of Energy Engineering, Faculty of Mechanical Engineering, Budapest University of Technology and Economics, Műegyetem rkp. 3., H-1111 Budapest, Hungary}, \orgaddress{\city{Budapest}, \country{Hungary}}}

\affil[4]{\orgname{Samara State Technical University, ul. Molodogvardeiskaya 244}, \orgaddress{\city{Samara}, \postcode{443100}, \country{Russia}}}

\affil[5]{\orgdiv{Mechanical Engineering Department, Faculty of Engineering}, \orgname{University of Zanjan}, \orgaddress{\city{Zanjan}, \postcode{45195-313}, \country{Iran}}}


\abstract{The accuracy of the classical heat conduction model, known as Fourier's law, is highly questioned, dealing with the micro and nanosystems and biological tissues. In other words, the results obtained from the classical equations deviate from the available experimental data. It means that the continuum heat diffusion equation is insufficient and inappropriate for modeling heat transport in these cases. There are several techniques for modeling non-Fourier heat conduction. In the present paper, we place our focus on the dual-phase-lag (DPL) approach. The DPL model, as a popular modification of Fourier's law, has already been utilized in numerous situations, such as simulating ultrafast laser heating and heat conduction in carbon nanotubes. There has been a sharp increase in research on non-Fourier heat conduction in recent years. Several studies have been performed in the fields of thermoelasticity, thermodynamics, transistor modeling, and bioheat transport. This review presents the most recent non-Fourier bioheat conduction works and the related thermodynamics background. The various mathematical tools, modeling different thermal therapies, and relevant criticisms and disputes are discussed. Finally, the novel and other possible studies are also presented to provide a better overview, and the roadmap to the future research and challenges ahead is drawn up.}

\keywords{Nonequilibrium thermodynamics, bioheat, non-Fourier heat conduction models, dual-phase-lag models}



\maketitle
\tableofcontents
\section{Introduction}
{{ 
		The delayed differential equation for non-Fourier heat conduction was first suggested by Tzou \cite{Tzo95a,Tzou1995,Tzou1995p2,Tzo97b,Tzou2010,Tzoubook}\footnote{Actually the expression `dual-phase-lag' appears in the review of Özisik and Tzou \cite{OziTzu94a}.} with the following constitutive equation for the heat flux: 		
		\begin{equation}
			\mathbf{q}(t+\tau_1,\mathbf{x}) = -\lambda \nabla T(t+\tau_2),
			\label{dpl_eq}\end{equation}
			called dual-phase-lag (DPL) equation.
		The content of the equation is that the temperature, $T$, and the heat flux, $\mathbf{q}$, are delayed by $\tau_1$ and $\tau_2$ periods. Here $\lambda$ denotes the thermal conductivity coefficient of the theory, which becomes identical with the Fourier one if the delay times are zero. In the present review, we aim to briefly present the non-Fourier models, and their relation with the DPL concept. Additionally, as there are numerous experimental results in the literature, we want to collect and summarize them as much as possible, in order to offer a clearer picture about the state-of-the-art understanding and achievements.

	Since 1994, the number of publications about non-Fourier heat conduction models involving time lag has been incredibly increased.  In fact, the development of novel models in order to modify the results of classical equations such as Fourier law with less computational cost and more simplicity has attracted notable attention. Recently, as the DPL model is intended to replace the Fourier law, that new approach has been tested by simulating the heat transport, e.g., in micro and nanoscales \cite{Vermeersch2008}, ultrafast processes \cite{Tzou2001,Chou2009}, living tissues \cite{Zhou2009}, and carbon nanotube \cite{Shiomi2006}.}
	 Figure~\ref{Fig1:stat} presents the number of publications after 1995 obtained by searching the keyword `dual-phase-lag' in Scopus and Web of Science databases. It is seen that the amount of research in the field of non-Fourier DPL heat conduction has experienced a sharp increase after about 2013, and that trend probably will continue. Ghazanfarian et al. \cite{Shomali2015} published a review paper and gathered various aspects of the lagging heat models, including mathematical models, solution methods, and applications.
	
	\begin{figure}
		\label{Fig1:stat}
		\centering
		\includegraphics[width=\columnwidth,height=0.7\columnwidth]{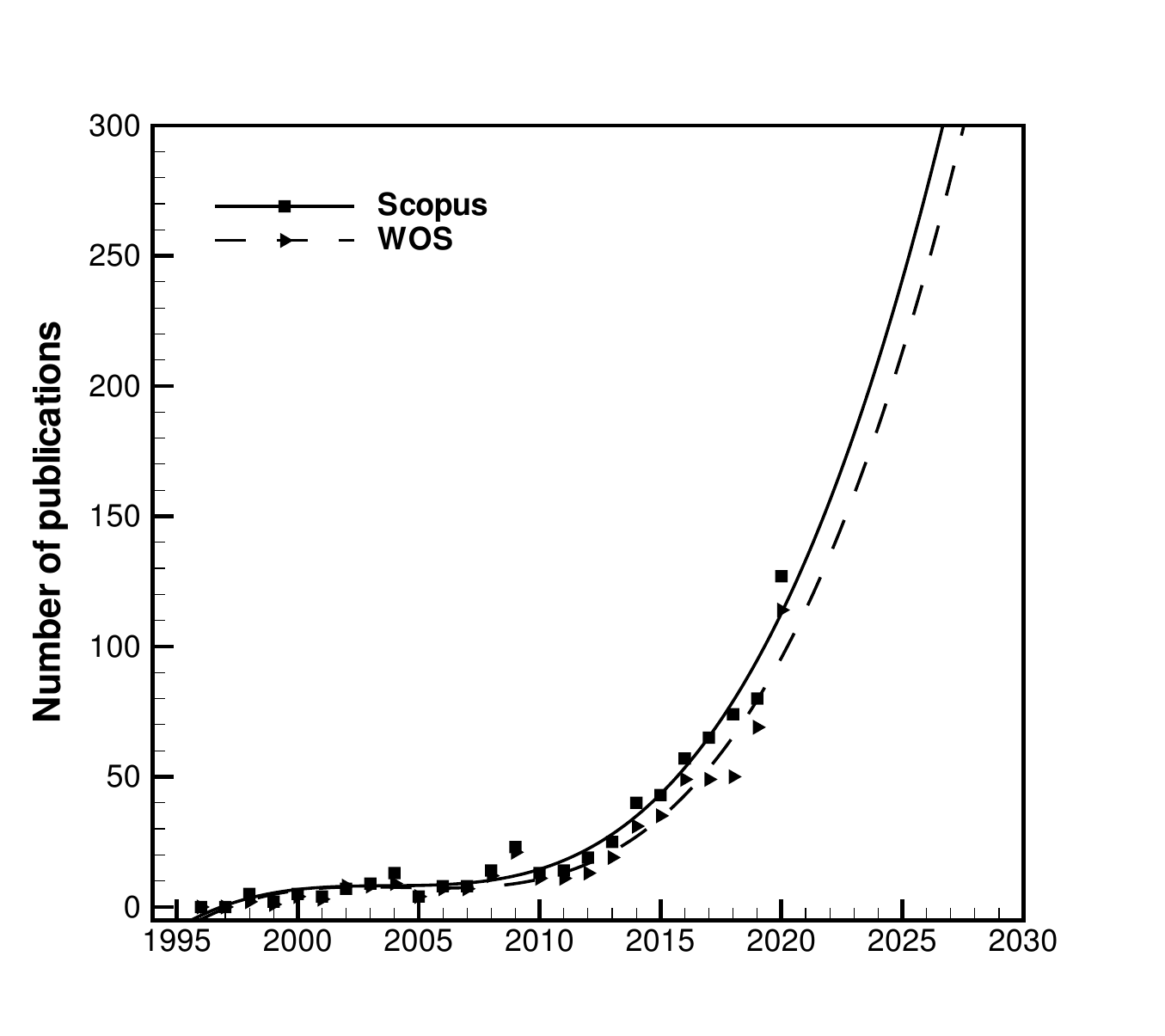}
		\caption{Statistics of publication history and future prediction for the number of publications, including research papers, book chapters, review papers in the last two decades and the future decade calculated, based on searching the keyword `dual-phase-lag' in Scopus and Web of Science databases.}
	\end{figure}

	{
		Despite the broad areas of applications the acceptance of the DPL model was controversial. The theoretical background, particularly the compatibility with basic physical principles, was criticized in several research papers. 
		In this review, we focus on two challenging topics of the field. First, the theoretical background is surveyed (Sec.~\ref{therm}), and the conditions of validity of the DPL concept is outlined. Then a review about bioheat modelling follows (Sec.~\ref{bio}), including the related solution methods. The review of developments of the lagging models in other fields such as thermoelasticity and microscale heat transport could be the topic of other review studies.
	}

%-------------------------------------------------------------------------------------------------------------------------
\section{Thermodynamics of heat conduction}
\label{therm}

%--------------------------------------------------------------------
\subsection{Theoretical studies}
%----------------------------------------
\subsubsection{Derivation of the Fourier equation}
\label{Fderi}
Heat conduction is the most frequent dissipative phenomenon, occurring in numerous engineering problems. Therefore, it is a starting point for all thermodynamic theories, particularly nonequilibrium thermodynamics. The distinctive property of heat conduction is that it does not have a reversible part, contrary to the dissipation-free mechanical systems. Fourier's law  is a prototype of the constitutive equations leading to a parabolic partial differential equation. Its simple classical derivation in irreversible thermodynamics is based on the second law of thermodynamics. We begin with the balance of internal energy. It is best written in a substantial form:
\begin{equation}
	\rho \dot e +  \nabla \cdot \mathbf{q} = 0,
	\label{ebal}\end{equation}
where $\rho$ is the mass density, $e$ is the specific internal energy, $\mathbf{q}$ is the heat flux, which is the conductive part of the current density of internal energy. The overdot denotes the substantial time derivative (i.e., $\frac{\textrm{d}}{\textrm{d}t} = \frac{\partial}{\partial t} + {\bf v}\cdot\nabla$) and $\nabla\cdot$ is the divergence operator. Here, internal energy is considered to be conserved with zero source term. In the classical local equilibrium situation the entropy depends only on the internal energy, and its derivative respect to the internal energy is the reciprocal temperature, this is true for the densities and for the specific entropy $s$, too,
$$
\frac{\textrm{d} s}{\textrm{d} e}(e) = \frac{1}{T}.
$$
Then, the temperature depends on the internal energy, and $T: e\rightarrow T(e)$ is the caloric equation of state. In order to close \eqref{ebal}, we need a relation  between the heat flux $\mathbf{q}$ and the temperature $T$. This relation is the so-called constitutive function that connects these field variables and is restricted by the second law of thermodynamics. As	 it is a constitutive relation, it reflects material properties and behavior. Its form can be determined using general principles by the second law, exploiting the requirement of non-negative entropy production $\Sigma$. The entropy inequality is conveniently written as
\begin{equation}
	\rho \dot s + \nabla\cdot \mathbf{J}=\Sigma \geq 0,
	\label{sbal}\end{equation}
in which $s$ is the specific entropy. The fundamental question is the form of the entropy flux $\mathbf{J}$. Eq.~\eqref{sbal} is a conditional inequality, where the balance of internal energy \eqref{ebal} must be considered as a constraint. According to Classical Irreversible Thermodynamics, both the entropy flux and entropy production can be determined with the following straightforward calculation,
\begin{equation}
	\rho \dot s(e) + \nabla\cdot \mathbf{J} = \frac{\rho\dot e }{T} + \nabla\cdot \mathbf{J} =
	\nabla\cdot \left(\mathbf{J} - \frac{\mathbf q}{T}\right) + \mathbf{q}\nabla\cdot \frac{1}{T}\geq 0.
\end{equation}
where the balance of internal energy \eqref{ebal} is also exploited.
The entropy flux is fixed by eliminating the first term in the last equality, that is,
$$
\mathbf{J}= \frac{\mathbf q}{T}.
$$
Thus the entropy production simplifies to
$$
\Sigma = \mathbf{q}\cdot\nabla \frac{1}{T}\geq 0.
$$
The simplest solution of this inequality is linear. For isotropic materials it is an isotropic expression, with a scalar coefficient, therefore
\begin{equation}
	\mathbf{q} = \lambda_T \nabla \frac{1}{T} = - \lambda \nabla T.
	\label{Flaw}\end{equation}
This is called Fourier's law, and  $\lambda = \frac{\lambda_T }{T^2}$ is the Fourier heat conduction coefficient with $\lambda_T$ being a thermodynamic - Onsagerian - conduction coefficient. The second law requires $\lambda_T >0$, otherwise, the entropy could decrease in closed systems as well. Substituting the caloric equation of state, $T(e)$, and the Fourier law into the balance of internal energy, one obtains the parabolic Fourier heat equation:
\begin{equation}
	\rho c \dot T - \nabla\cdot (\lambda \nabla T) =0,
	\label{Feq}
\end{equation}
in which $c = \frac{\textrm{d} e}{\textrm{d}T}$ is the specific heat. In case of constant coefficients $\rho, c,\lambda$, { particular initial-boundary value problems have unique solution and also are well-posed \cite{Dib95b}.} 
It is remarkable that in the case of Neumann boundary conditions, from the zero temperature gradient follows that the heat flux is zero at the boundary, $\mathbf{q}_{bou}=0$.

\subsubsection{Application limits of the Fourier equation}
The above derivation of the Fourier equation is based on the principle of local thermodynamic equilibrium and the continuum hypothesis. The state of the system, in this case, can be described by local thermodynamic potentials that depend only on the spatial variable and time through thermodynamic parameters. Accepting the local equilibrium hypothesis is possible only if the rate of change in the system macroparameters due to external influences is much less than the rate of system relaxation to local equilibrium \cite{Sobolev1991,Sobolev1997}. {In other words, one can introduce
\begin{equation}
	\nonumber
	v_1=L/t_0; \; \; v_2=l/\tau
\end{equation} 
two characteristic speeds, where $v_1$ is the linear rate of parameter change and $v_2$ is the rate of disturbance propagation. The first one is caused by asymmetric boundary conditions (e.g., isothermal displacement rate), and characterized by the system size $L$ and the time ($t_0$) of the process until equilibrium is established.
Regarding the second one, $v_2$ is the rate of disturbance propagation, which is characteristic for the internal system, independently of the boundary conditions. In a statistical context, $l$ is the characteristic scale of the microstructure, and $\tau$ is the relaxation time characterizing the free path time of the microparticles.
Hence, the infinite rate of disturbance propagation described by the parabolic transport equations is related to the fact that the relaxation time is assumed to be zero. Consequently, the disturbance propagates instantaneously.} To eliminate this disadvantage in parabolic equations, it is necessary to develop a mathematical theory to describe the transport processes occurring under locally nonequilibrium conditions. Various methods can be used to describe these processes, such as thermodynamic, molecular-kinetic, and phenomenological, as well as those based on the random walk theory and the thermal memory concept \cite{Sobolev1991,Sobolev1997,Kudinov2014, Kudinov2020}. In most of these methods, the conditions of local thermodynamic equilibrium and the continuum hypothesis are violated, the molecular-atomic structure of matter is taken into account, resulting in a delay in transport time.

Biomaterials, active matter are not in local thermal equilibrium, therefore classical Fourier theory cannot be applied without any further ado. However, their internal structure is complex at the mesoscopic level, therefore direct microscopic derivations cannot be applied.
%--------------------------------------------------------------------
\subsubsection{Paradoxes: relation of memory and speed}
There is one crucial assessment of the parabolic Fourier heat equation from a physical point of view. This is the problem of infinite signal propagation speed, the so-called \emph{heat conduction paradox}. This is an inevitable mathematical fact, but it is not a sufficient reason to look for non-parabolic models. The primary implicit motivation behind it is an expectation attributed to special relativity. However, this expectation is completely unfounded due to several reasons.
\begin{enumerate}
	\item The practical and theoretical range of validity of the Fourier theory determines finite signal propagation speeds, and, for usual materials, is far below the speed of light. The practical limit is the sensitivity of measurements, and the theoretical limit is the validity range of the theory being a continuum \cite{Wey67a,Fic92a,KosLiu00a}. For instance, in the case of fluids, the temperature field must not vary too rapidly over the average mean free path, $\langle d\rangle$, and average time between collisions, $\langle t\rangle$, of microscopic constituents (particles or elementary excitations):
	$$
	\left[\frac{1}{T}\frac{\partial T}{\partial x} \right] << \frac{1}{\langle d\rangle}, \qquad
	\left[\frac{1}{T}\frac{\partial T}{\partial t} \right] << \frac{1}{\langle t\rangle}.
	$$
	For water in room temperature the first condition gives the speed limit $v_{max} = \frac{\lambda}{\rho c \langle d\rangle} \approx 14 m/s$.
	\item 
	For a symmetric hyperbolic partial differential equation with a finite speed of signal propagation on nonrelativistic space-time, the propagation speed is a material property and has nothing to do with special relativity. It can be larger or smaller than the speed of light, depending on the parameters in the differential equation. More importantly, the underlying theory itself does not restrict the propagation speed from the above {\cite{KosLiu00a,VanBir08a}}.
\end{enumerate}

However, according to the first point, a particular solution of a parabolic equation with given initial and boundary conditions determines the domain of applicability. For instance, the gradient in the above inequality is time and space-dependent. Therefore when the condition is violated, we are beyond the range of validity of the Fourier heat equation. Consequently, a new approach is necessary. This must be a theory that improves the problematic aspects such as limiting the too high theoretical propagation speed of the continuum limit. We have arrived at the starting point, but with a clarified motivation in the background: to improve the parabolic theory's performance, one should consider a memory, a delay in the response of the physical system. { From a technical point of view, memory effects are represented in three different forms: a strong time nonlocality with memory functionals, weak time nonlocality with time derivative expansions, and a directly delayed mathematical description. These three realizations are related to each other, as explained in the next section.}

%--------------------------------------------------------------------
\subsubsection{Various memory types}
A particular form of memory is inertia. In systems with inertia, the motion does not stop when the force diminishes. { From a mathematical point of view, the representation of inertia requires additional state variables, e.g., momentum in mechanics and heat flux in heat conduction. In heat conduction, the time derivative of the heat flux appears in the constitutive equation, and one obtains the so-called Maxwell-Cattaneo-Vernotte (MCV) equation}:
\begin{equation}
\tau \dot {\mathbf q}+\mathbf{q} = \lambda_T \nabla \frac{1}{T} = - \lambda \nabla T.
\label{MCV}\end{equation}
Here, $\tau$ is the relaxation time, expressing inertial effects in heat conduction. This equation was suggested intuitively by Maxwell \cite{Max867a}, Cattaneo \cite{Cat48a} and Vernotte \cite{Ver58a1}. The elimination of the heat flux leads to a telegraph-type, damped wave propagation equation for heat conduction. The MCV equation can be formulated equivalently in the framework of Gurtin-Pipkin theory, with the following convolution integral,
$$
\mathbf{q}(t,\mathbf{x}) = - \int_{-\infty}^t \kappa(t-s) \nabla T(s,\mathbf{x}) \textrm{d}s,
$$
where the memory kernel is $\kappa(s) =
\frac{\lambda}{\tau} e^{-\frac{s}{\tau}}$ \cite{ColGur67a,GurPip68a,FabLaz14a}. This is the previously mentioned strong time nonlocality utilizing a memory functional. Analogously, one may imagine that there is a time difference, a delay, a lag between the temperature and heat. That is, the Fourier law is modified as
$$
\mathbf{q}(t+\tau_1,\mathbf{x}) = -\lambda \nabla T(t+\tau_2),
$$
{suggested by Tzou in the dual-phase-lag theory, as it was already mentioned in the Introduction \eqref{dpl_eq}}. Thus one obtains a delay partial differential equation, for which by applying the first-order Taylor series expansion, it reproduces the MCV equation with $\tau= \tau_1-\tau_2$.

{We have seen three different ways to consider memory effects, the inertial terms, the memory functionals, and the time-lag modifications.
They are seemingly more or less equivalent in the case of temperature representation, i.e., when only the temperature $T$ is used as a field variable after eliminating the heat flux, with the energy balance Eq.~\eqref{ebal}. Although different approaches can lead to the same PDE for the temperature, they still differ in the constitutive part, which restricts how the initial and boundary conditions can be defined. Therefore, they stand for a different physical meaning as well. The differences come into the light when one attempts to solve these models with time-dependent heat flux boundary conditions numerically and analytically. Also, their application to practical problems, especially with nonlinearities can highlight the essential differences, both for mathematical and physical aspects. The aforementioned constitutive equations restrict how the boundary conditions can be formalized.} This is discussed later in Section 2.4. Moreover, it is noticeable that these approaches can be consistent on the level of memory extensions. However, when one must go beyond the MCV equation and a simple memory effect is not enough, the different approaches are more likely to deviate from each other.
In order to understand their possible limitations, one should analyze the principles behind the derivation of the Fourier law.

%--------------------------------------------------------------------
\subsubsection{Fundamental principles: space-time and the second law}
We have already mentioned the mathematical requirements, in particular, well-posedness. From a physical point of view, there are two further fundamental aspects: heat conduction is a continuum phenomenon, therefore requires a space-time representation. On the other hand, it is a dissipative phenomenon as well, thus it must be compatible with the second law of thermodynamics. One can observe the appearance of these aspects in the heuristic derivation of Fourier's law from classical irreversible thermodynamics in Section~\ref{Fderi}.

First of all, one should realize that the substantial time derivatives, the conductive current densities in the internal energy, and the entropy balances are tools to attach the reference frame to the flow of the material. Substantial time derivatives are equal with partial time derivatives in the case of rigid heat conductors when the heat-conducting substance does not move. Also, the internal energy itself is the difference of total and kinetic energies based on the concept of comoving energy, as it is clear from the transformation properties \cite{VanEta19a}. Fourier's law is objective, but the generalizations with time derivatives such as the MCV equation, the material frame indifference is open and discussed question \cite{Mul72a,ChrJor05a,Chr09a}. It is crucial when memory effects are mixed with spatial nonlocality and mechanical couplings \cite{Van17a,Van20a1}. The dual time lag approach is not space-time compatible, e.g., violates the homogeneity of time, one of the basic requirements of any space-time. Also, the separation of time from space, as appears in both the memory integrals and in time lag equations, cannot survive relativistic generalizations. A covariant, objective representation is inevitable since the Minkowski form is absolute, but space and time are reference frame-dependent. Let us emphasize here that it is an unavoidable aspect also in a nonrelativistic framework, where time is absolute instead of the light speed \cite{Van17a,VanEta17a}.

The second fundamental aspect is the second law and the consequent entropy balance analysis. A theory that violates the second law is not acceptable. Therefore, the entropy function and the entropy inequality are good starting points for generalizations. The direct time lag approach neglects this requirement, consequently, encounters conceptual, mathematical and stability { problems \cite{DreEta09a,FabLaz14a,FabFra14a,RukSam13a,Ruk14a,FabEtal16,Ruk17a,ChiEta17a,KovVan18a1}. }According to these results DPL model can be compatible with the second law only if a differential version is considered and in case of particular conditions for the parameters (see also \cite{AskEta18a}). Those conditions look compatible with the principle of fading memory. In the case of memory kernels, this principle is a particular form of the second law. However, it is not constructive, especially for spatial nonlocality, in which case, extra space derivatives and spatial scaling properties appear, e.g., at the nano level.

%--------------------------------------------------------------------
\subsubsection{Theories of heat conduction beyond Fourier}
Any method or theoretical framework requires benchmarks, tests of performance. Firstly, they must be able to \emph{predict experimentally observable effects}. The various empirical extensions of the Fourier equation, see, e.g., \cite{JosPre89a,JosPre90a}, provide a reasonable test area. Their mathematical and physical consistency are different, and also, the experimental observations may be confusing, but a uniform theoretical background could clarify the conditions of their validity.

Secondly, a macroscopic theoretical framework must be compatible with the kinetic theory of gases. Rarefied gases have the best understood microscopic composition among continua. The macroscopic equations from the moment series expansion of kinetic theory are instructive in this respect. It is remarkable that kinetic theory alone, with its specific assumptions about the structure of materials, cannot substitute a thermodynamic treatment. The experimentally observed validity of the macroscopic constitutive equations, in particular the Fourier's law, does not depend on the particular microscopic structure. The second law and the space-time requirements of nonequilibrium thermodynamics are universal, independent of material composition and properties. Therefore their consequences are universal as well. Now, considering the previously mentioned space-time and second law-related requirements, the inertial and memory effects are best introduced by the extension of the thermodynamic state space with new fields, called internal variables. In the following, we show the simplest possible example, Fourier's theory is extended by a single internal variable. Here we do not use any sophisticated space-time concepts beyond the already introduced ones except that entropy flux will be considered as a constitutive quantity restricted by the second law. For that purpose, it is convenient to use Ny\'\i{}ri multipliers \cite{Nyi91a1}.

Several theoretical frameworks extend the classical thermodynamic state space with additional fields. The theories where the second law plays a crucial role are Rational Extended Thermodynamics (RET) \cite{MulRug98b,DreStr93a}, Extended Irreversible Thermodynamics (EIT) \cite{Gya77a,JouAta92b,LebEta08b,LebEta11a,CasJou03a}, Non-Equilibrium Thermodynamics with Internal Variables (TIV) \cite{Ver83a,Nyi91a1,VanFul12a,KovVan15a} and Rational Thermodynamics (RT) \cite{GreNag91a,FabMor03a,Mar17a,CapEta21a}. The relation of these theories to the previously mentioned requirements (well-posedness, second law, and space-time) and benchmarks (experimental, kinetic theory compatibility) are very different, and also, their applicability is not the same (see also \cite{Van20a}). RET, EIT, and TIV are constructive. RT is capable of checking the second law compatibility of the suggested constitutive equations. RET, EIT, and TIV pay attention to the compatibility with the kinetic theory of gases, where RET is the most strictly compatible, TIV is the least compatible, and EIT is situated between them.
Furthermore, the symmetric hyperbolic structure of the equations is essential. In RET, the construction method exploits the thermodynamic potentials to get symmetric hyperbolic evolution equations. EIT is open for the possibility, while in TIV the second law compatibility is the most important, and the entropy production inequality is the basis of construction. On the contrary, these approaches can be compatible with each other under certain conditions.

%--------------------------------------------------------------------
\subsubsection{Memory and spatial nonlocality}
In a complete theory, the inertial effects cannot be separated from a gradient extension. Memory effects require nonlocal modification, due to the space-time representation. Let us demonstrate the interplay of time derivatives, gradient terms and the differences of the previously mentioned theories with the example of TIV using a single internal variable. The thermodynamic state space is spanned by the internal energy $e$, and a vectorial internal variable $\boldsymbol{\boldsymbol{\xi}}$. Therefore, the specific entropy depends on these variables $s(e,\boldsymbol{\xi})$, and the thermostatic relations, the potential properties of the thermodynamic state space are given in the traditional form of differentials with the following Gibbs relation,
\begin{equation}
	{\rm d}e = T{\rm d}s - T {\mathbf Z}\cdot {\rm d}\boldsymbol{\xi},
	\label{Grel}
\end{equation}
where
\begin{equation}
	\nonumber
	\frac{\partial s}{\partial e}\vert_{\boldsymbol{\xi}} = \frac{1}{T}, \ \ \ \ \ \ \
	\frac{\partial s}{\partial \boldsymbol{\xi}}\vert_{e} =  \mathbf Z.
\end{equation}
The Gibbs relation is a convenient form to introduce the partial derivatives of the thermodynamic potentials \cite{BerVan17b}. It is also important to consider the entropy flux as a constitutive quantity. In this respect, Ny\'\i{}ri (or current) multipliers provide the necessary flexibility. In our case $\mathbf J =\mathbf b\cdot\mathbf q$, that is we assume that there is no entropy flux if the heat flux is zero and the second order tensor $\mathbf b$ is the Ny\'\i{}ri multiplier. It is a constitutive quantity, a function, whose form is restricted by the second law. With these assumptions the entropy production can be calculated as
\begin{eqnarray}
	\rho \dot s(e,\xi)+\nabla\cdot {\mathbf J} = \frac{\rho \dot e}{T} + \rho{\mathbf Z}\cdot \boldsymbol{\dot\xi} + \nabla\cdot{\mathbf J} = \nonumber\\
	-\frac{\nabla\cdot{\mathbf q}}{T} + \rho {\mathbf Z}\cdot\boldsymbol{\dot\xi} + (\nabla\cdot {\mathbf b})\cdot {\mathbf q} + {\mathbf b}:\nabla {\mathbf q} = \nonumber\\
	(\nabla\cdot {\mathbf b})\cdot {\mathbf q} + \rho {\mathbf Z}\cdot\boldsymbol{\dot\xi} +\left({\mathbf b} - \frac{1}{T}{\mathbf I}\right):\nabla {\mathbf q}\geq 0,
	\label{hm}\end{eqnarray}
in which double dot denotes the trace of product of two second order tensors, $\mathbf{A}:\mathbf{B} = Tr(\mathbf{A}\cdot\mathbf{B})$, and  $\mathbf I$ is the identity tensor. The constitutive quantities are $\mathbf b$, $\mathbf q$ and the evolution equation of the internal variable
$$
\boldsymbol{\dot\xi} = \boldsymbol{\Xi}(e,\boldsymbol{\xi}).
$$
That is an important feature of internal variable theories, the right hand side is to be determined constitutively, including the differential operator, $\boldsymbol{\Xi}$. Therefore, the last line of Eqs.~\eqref{hm} is a quadratic expression of thermodynamic fluxes and forces, where the thermodynamic fluxes are the constitutive multipliers in the quadratic expression of the entropy production  (see Table 1). This assumption is in complete agreement with Onsager's idea \cite{GroMaz62b}.
\begin{center}
	\begin{tabular}{c|c|c|c}
		& Classical thermal & Extended thermal & Internal \\ \hline
		Fluxes & $ {\mathbf q}$ &
		${\mathbf b} -\frac{1}{T}{\mathbf  I} $ &
		$  \boldsymbol{\dot\xi}$\\ \hline
		Forces &$\nabla\cdot{\mathbf b}$ &
		$\nabla {\mathbf q}$ &
		${\mathbf Z} $\\
	\end{tabular}\\
	\vskip .21cm
	{Table 1. Thermodynamic fluxes and forces}\end{center}

The classical linear solution of the inequality for isotropic materials is the following:
\begin{eqnarray}
	{\mathbf q} &=& l_1 \nabla \cdot {\mathbf b} + l_{12} {\mathbf Z}, \label{o1}\\
	\rho\boldsymbol{\dot\xi} &=& l_{21} \nabla \cdot {\mathbf b} + l_{2} {\mathbf Z}, \label{o2}\\
	{\mathbf b} - \frac{1}{T}{\mathbf I} &=& k_1 (\nabla{\mathbf q})^{s0} + k_2 \nabla\cdot{\mathbf q}{\mathbf I} +k_3 (\nabla{\mathbf q})^{as},
	\label{o3}\end{eqnarray}
where $(\nabla{\mathbf q})^{s0} = \frac{1}{2}(\nabla{\mathbf q}+\nabla{\mathbf q}^*)/2- \nabla\cdot{\mathbf q}{\mathbf I}/3 )$ is the symmetric traceless part of the second order tensor, the so-called deviatoric part, where $*$ denotes its transpose. Also, $(\nabla{\mathbf q})^{as} = \frac{1}{2}(\nabla{\mathbf q}-\nabla{\mathbf q}^*)/2$ is the antisymmetric part. %The middle term at the right hand side of~\eqref{o3} is called the spherical part.
The coefficients are not arbitrary, they are restricted by the following inequalities:
\begin{equation}
	l_1\geq 0, \quad l_2 \geq 0, \quad l_1l_2 - \frac{(l_{12}+l_{21})^2}{4} \geq 0, \quad k_1,k_2,k_3\geq 0.
	\label{slineq}\end{equation}
This is the consequence of the second law inequality, the requirement of non-negative entropy production. Moreover, these relations are essential when the coefficients are state-dependent, e.g., the thermal conductivity depends on the temperature. That dependence affects the other parameters, and therefore, the treatment of nonlinear problems are not starightforward and could be model-dependent. In regard to the DPL equation, these properties are missing.

Then one can easily eliminate the Ny\'\i{}ri multiplier, $\mathbf b$, using~\eqref{o3}. The internal variable field, $\boldsymbol{\xi}$, can be eliminated as well, assuming a particular equation of state for $\mathbf Z$. If the nonequilibrium field contributes quadratically to the internal energy, that is, $s(e,\boldsymbol{\xi})=\hat s(e-m\boldsymbol{\xi}^2/2)$, one obtains $\mathbf Z= -m\boldsymbol{\xi}$, and in case of constant coefficients a straightforward calculations leads to the following  evolution equation of heat flux
\begin{eqnarray}
	\label{eq.13}
	\tau \dot {\mathbf q}+ {\mathbf q}=\lambda \nabla\frac{1}{T}+
	\hat\lambda \frac{\rm d}{\textrm{d} t}\left(\nabla\frac{1}{T}\right) +
	\lambda\kappa_1 \nabla\cdot\nabla{\mathbf q} +\nonumber\\
	\hat\lambda \kappa_1\frac{\rm d}{\textrm{d} t}\left(\nabla\cdot\nabla {\mathbf q}\right) +
	\lambda\kappa_2 \Delta{\mathbf q} +
	\hat\lambda \kappa_2\frac{\rm d}{\textrm{d} t}\left(\Delta {\mathbf q}\right).
	\label{genhq}\end{eqnarray}
The time and space derivatives are commutative in rigid heat conductors. The parameters in Eq.~\eqref{eq.13} are defined as $\tau = \frac{\rho}{m l_2}$, $
\lambda=\frac{l_1l_2-l_{12}l_{21}}{l_2}$, $\hat\lambda=\frac{\rho l_1}{m l_2}$,
$\kappa_1=\frac{k_1+2k_2+3k_3}{6}$ and $\kappa_2=\frac{k_1\! -\! k_3}{2}$. Remarkable, that all these coefficients are nonnegative except the last one. One can see that Eq.~\eqref{genhq} is a generalization of both the Fourier's law, Eq.~\eqref{Flaw}, and the MCV equation, Eq.~\eqref{MCV}, and incorporates several other ones \cite{VanFul12a}, seemingly including the Jeffreys-type and the  Guyer-Krumhansl equations. It is also remarkable that the coefficients are not independent and the second law, the inequalities in~\eqref{slineq} require $\tau_0\geq 0$, $\lambda\geq 0$, $\hat \lambda\geq 0$ and $\kappa_1>0$.

Noticeably, the Guyer-Krumhansl equation is obtained when $l_1=0$. The heat flux $\mathbf q$ is proportional to the internal variable $\boldsymbol{\xi}$, according to~\eqref{o1}, because $\mathbf Z= -m\boldsymbol{\xi}$. Therefore, a rescaling of the internal variable leads to the Extended Thermodynamic theories with the heat flux as an independent field, and the Ny\'\i{}ri multiplier $\mathbf b$ as the flux of the heat flux in the balance form~\eqref{o2}. It is also remarkable that in this case $\hat \lambda=0$ and $l_{12}l_{21} \leq 0$, due to the third inequality of~\eqref{slineq},  thus still $\lambda \geq 0$.

By keeping the terms in the left and right sides of the Eq.~\eqref{eq.13}, the resulting equation for temperature will be parabolic. For the convenience, let us write Eq.~\eqref{eq.13} in one spatial dimension and in the form in which all derivatives are kept, the coefficients are written according to the dimensions fulfilment condition,
{
\begin{equation}
	\label{igor1}
	q +\tau \frac{\partial q}{\partial t} =
	-\lambda_F \frac{\partial T}{\partial x}-\lambda_F \tau \frac{\partial^2 T}{\partial x \partial t}+l^2 \frac{\partial^2 q}{\partial x^2}+l^2 \tau \frac{\partial^3 q}{\partial x^2 \partial t},
\end{equation}}
where $\lambda_F$ is the Fourier thermal conductivity coefficient; $\tau$ is the heat flux and temperature gradient relaxation coefficient; and $l$ is the free path length of the {\em energy carriers}, which can be related to the characteristic size of material heterogeneities. Note that  $\lambda_F = \lambda/T^2$ was assumed to be constant in the derivation and also in Eq.~\eqref{igor1}, the relaxation coefficients for the heat flux and temperature gradient and heat flux derivatives are assumed to be the same. The fact that the relaxation coefficient condition does not coincide with those given in Eq.~\eqref{eq.13}, is an experimental question. More importantly, they are positive (see \eqref{slineq}) and the dimensions of the terms with derivatives in Eq.~\eqref{igor1} that coincide with the one in Eq.~\eqref{eq.13} remains the same. By substituting Eq. \eqref{igor1} into the internal energy balance equation, the following equation for temperature
is obtained,
\begin{equation}
	\label{igor2}
	\frac{\partial T}{\partial t}+\tau\frac{\partial^2 T}{\partial t^2}=\alpha \frac{\partial^2 T}{\partial x^2}+(\alpha\tau+l^2)\frac{\partial^3 T}{\partial x^2 \partial t}+l^2\tau\frac{\partial^4 T}{\partial x^3 \partial t},
\end{equation}
with $\alpha=\lambda/(\rho c)$ thermal diffusivity. It is apparent that Eq.~\eqref{igor2} is parabolic because it contains two terms with mixed derivatives and lacks a term with a third time derivative which can occur only when  Eq.~\eqref{igor1} contains the $\tau^2\partial^2 q/\partial t^2$ term. One hyperbolic equation for heat conduction can be obtained using { ballistic conductive constitutive equation of \cite{KovVan15a} or the second-order heat flux in} \cite{Sobolev1997},
\begin{equation}
	\label{igor3}
	q=-\lambda\frac{\partial T}{\partial x}-\lambda \tau_T \frac{\partial^2 T}{\partial x \partial t}-(\tau_T+\tau_q)\frac{\partial q}{\partial t}-\tau_T\tau_q \frac{\partial^2 q}{\partial t^2}+l^2 \frac{\partial^2 q}{\partial x^2},
\end{equation}
where $\tau_T$ and $\tau_q$ are the relaxation times for the heat flux and temperature gradient. By substituting Eq. \eqref{igor3} into the heat balance equation, the following equation is derived
\begin{equation}
	\label{igor4}
	\frac{\partial T}{\partial t}+(\tau_T+\tau_q)\frac{\partial^2 T}{\partial t^2}+\tau_T\tau_q \frac{\partial^3 T}{\partial t^3}=\alpha\frac{\partial^2 T}{\partial x^2}+(\alpha\tau_q+l^2)\frac{\partial^3 T}{\partial x^2 \partial t}.
\end{equation}
%This is also the so-called T-wave equation that corresponds to the second-order expansion of Eq.~\eqref{dpl_eq}.
This is also the so-called T-wave equation that can be connected to the second-order expansion of Eq.~\eqref{dpl_eq}. However, the coefficients are interpreted differently: while the Taylor series expansion of Eq.~\eqref{dpl_eq} would result in coefficients with time delays, Eq.~\eqref{igor4} corresponds to the ballistic-diffusive model. The structure of the equation is the same in both cases. Please note that the higher-order approximations of a thermodynamic theory are thermodynamically consistent, too. Moreover, heat flux is not a scalar quantity, and higher-order tensorial properties of heat conduction are not trivial, neither in the case of isotropic materials \cite{FamEta21a}.

It is clear that Eq.~\eqref{igor4} is hyperbolic. The exact analytical calculations present that depending on the values $\tau_q$, $\tau_T$, $l$, the equation can describe diffusion as well as wave and ballistic heat exchange modes. The last two can be observed in nanofilms with a thickness comparable to the free path length of the microparticles and also in superfluid He-4.

In RET and EIT, the thermodynamic state variable is the heat flux from the beginning and has a balance form evolution such as~\eqref{o2}. These are postulated and not derived. Consequently, one cannot get the Jeffreys-type equation, the second term on the right hand side is missing. In RET, the symmetric hyperbolic structure and the particular form of the source terms reduce the number of independent material parameters compared to EIT or TIV. Further differences are analyzed in \cite{CimEta14a,Jou20a,KovEta20a}.

The collection of thermodynamic theories is not complete without mentioning GENERIC \cite{Ott05b}, the conservation-dissipation formalism \cite{Yon20a}, and the Symmetric Hyperbolic and Thermodynamically Compatible (SHTC) equations \cite{RomEta20a}. The symmetric hyperbolicity, i.e., the Hamiltonian structure, acts as a basic requirement. In the present review, we {focus on} theories of heat conduction with experimental predictions or experimental comparisons, and in this respect, the aforementioned theories are poor.

%--------------------------------------------------------------------
\subsubsection{Further memories, inertia of the inertial terms}
A straightforward generalization is when one looks for an inertial effect regarding the nonequilibrium field, the memory of the memory effect. Practically, this means looking for a theory with further, higher-order time derivatives in the evolution equation beyond the MCV term, for instance, the one needed to describe ballistic propagation. Ballistic propagation was predicted theoretically with kinetic theory for rarefied gases.

In phonon hydrodynamics, it is attributed to the free propagation of the particles when the average mean free path is larger than the characteristic length of the system. For phonons, for the quanta of lattice oscillations, free propagation means propagation with the speed of sound. From a continuum point of view, it is the limit of ideal elasticity. Ballistic propagation with a speed of sound is a thermoelastic effect in a continuum framework. Therefore, mechanics and thermodynamics should be carefully analyzed and coupled in any continuum framework when modeling ballistic propagation \cite{FriCim95,FriCim96, FriCim98, BallEtal20}. In RET and EIT, where the structure of the evolution equations follows from the momentum series expansion of kinetic theory, the highest speed of propagation approaches the speed of sound, which is characteristic for ballistic propagation, when the number of moments is high \cite{MulRug98b,DreStr93a}.

The direct coupling of an integral model of collisionless propagation of ballistic phonons to an MCV-like diffusive continuum theory is a separate theory, called ballistic-diffusive \cite{Che01a,Che02a}. { A third approach to ballistic propagation is possible with a direct continuum point of view. Introducing a second tensorial internal variable in addition to the vectorial one in Section 2.1.7. \cite{KovVan15a} gives a clear connection to thermal and mechanical interaction, including the propagation of thermal disturbances with the speed of sound and the thermal expansion effect \cite{BerBer13a,BerEng13a,BerVan16a,BerVan17b,Ber19a}. This leads to Eq.~\eqref{igor4} as a special case.} It is also remarkable that the integro-differential equations of the ballistic-diffusive theory can be substituted with a two-component differential model, coupling an MCV-type propagation to a Guyer-Krumhansl-type propagation of the diffusive and ballistic heat carriers \cite{LebEta11a}. The conceptual RET and TIV-based models are best compared in case of rarefied gases, which is analogous to ballistic heat propagation. Although the experiments are less technical, their evaluation demands to investigate other, not straightforward aspects such as mass density dependence of material properties \cite{AriEtal12c,RugSug15b,AriEta20a,KovEta20a}. Continuum mechanics and heat conduction are strongly related. Ballistic heat conduction is the benchmark where the continuum and kinetic theories can be compared and tested.

%-------------------------------------
{\color{black}\subsection{ Heat conduction experiments}
	
The previous considerations, and particularly Eq.~\eqref{genhq}, are independent of any assumptions regarding the structure of the material. It can either be a biological tissue or a piece of rock. The compatibility with kinetic theory and special relativity are mentioned only as benchmarks. The assumptions are the second law, energy conservation, and material symmetries. The internal variable is a general method of quantitative characterization of the deviation from the local equilibrium: it can express different structural effects, like local multi-temperature nonequilibrium, material heterogeneity, delay in heat conduction, and so on. We have seen that it can be eliminated for isotropic materials, and we obtain the corresponding modifications of the Fourier law.

Other approaches, such as DPL or fractional derivative models, have a weaker theoretical background; however, they can be suitable in particular cases. How much are theories helpful in the interpretation of observations? The experimental tests are essential in distinguishing between the various theories because new observations may require more insight and a more detailed understanding of the particular phenomenon. Nevertheless, one must design the experiments in a way to keep the focus on the material behavior. For instance, separate the effect of source terms from the constitutive model. This way, it becomes possible to distinguish between various phenomena, and it significantly aids the proper interpretation of the observed data. The subsequent experimental studies summarize the phenomena which can be purely understood and modeled with generalized models.}

%-------------------------------------
\subsubsection{Experimental studies}

{The existence of numerous theoretical models requires a reliable selection criterion and validation method in which the experiments stand as the most substantial feedback for the theories. The previous decades were a fruitful era in that respect, starting from the predictions of low-temperature phenomena, which were followed by different room temperature observations.} Now, let us summarize the known heat conduction modes:
\begin{itemize}
\item diffusive: this is modeled by the classical Fourier's law,
\item second sound: this is a damped wave propagation of heat, which falls beyond Fourier's law and thus requires its generalization including memory effects,
\item ballistic: a thermo-mechanical phenomenon in which heat is conducted with the speed of sound, this is the fastest mode of heat conduction,
\item over-diffusive: although it is also a diffusive mode, its observation requires heterogeneous material structure, which cannot be modeled with the Fourier equation.
\end{itemize}

{In the following, we go along the experimental background, especially about the interpretation of measured temperature signals. Usually, the experiments are performed using a laser flash or heat pulse-type experimental apparatus. The sample is thermally excited on its front side by a short heat pulse, and the temperature is recorded on the other side.} This is a standard method to measure the thermal diffusivity and is also suitable due to the wide range of time scales that can be investigated \cite{ParEtal61}. There are geometrical restrictions in order to keep the heat conduction phenomenon on a one-dimensional level as much as possible. However, this is true only for uniform excitation on the surface and isotropic materials. The history of non-Fourier heat conduction has begun with the theoretical predictions of Onsager, Tisza, and Landau. While Onsager argued about probable microstructural reasons for deviation from Fourier's law \cite{Onsager31I}, such as for heterogeneous materials, Tisza and Landau - in some sense, analogously - considered a two-fluid model for superfluid states of helium \cite{Tisza38, Lan41}. In between, the common ground lies in the existence of parallel conduction channels inside the material, which might occur both in heterogeneous materials and at low-temperature states.

%--------------------------------------------------------------------
\subsubsection{Low-temperature heat conduction}
Most frequently, the second sound was observed in low-temperature situations \cite{NarDyn72a, NarDyn75, Naretal75}. This was first measured by Peshkov in 1944, using superfluid helium II \cite{Pesh44}. The superfluid state still has great importance from many aspects such as turbulence \cite{MongEtal18, SalJou20, SciEtal19, BewEtal06}, phase diagrams \cite{KimCha04, VollWolf13b}, viscosity properties \cite{London54b}, and how they affect the outcome of a thermal process \cite{HohMar65, Putt74, Dres82b, Dres84b}. Several questions arise about the propagation speed of the second sound. Its modeling requires, e.g., the Maxwell-Cattaneo-Vernotte equation \cite{Max867a,Cat48a, Ver58a1}, the first - hyperbolic - extension of Fourier's law. The predicted characteristic wave speed is $v=\sqrt{\alpha/\tau}$ in which $\alpha$ is the thermal diffusivity, and $\tau$ is the relaxation time. Their ratio characterizes the observed wave, and therefore, the thermophysical properties. However, as many authors pointed out \cite{LaneEtal46, MauHer49, WaWi51, Pell49}, the propagation speed is highly nonlinear. Consequently, all the thermal parameters must depend on the temperature \cite{MasRom17}. Depending on the material, that change could cover an order of magnitude, especially close to $0$ K \cite{AtkOsb50, JacWal71, JacWalMcN70}. 

{ Notably, including such nonlinear behavior in the heat conduction model is possible only with an established thermodynamic background since the coefficients are not independent of each other. The Onsagerian relations connect them, the temperature dependence of thermal conductivity influences the other parameters \cite{KovRog20}. This is entirely missing from the DPL model.} It has far-reaching consequences. It turned out that the temperature dependence of relaxation time necessarily implies the temperature dependence of mass density \cite{KovRog20}, and thus the physical interpretation demands a complete thermo-mechanical framework. Hence we arrive at ballistic propagation, which is an elastic wave carrying heat. From a continuum point of view, it is induced by thermal expansion \cite{FriCim95, FriCim96, FriCim98, BallEtal20}, and always has the speed of sound \cite{McN74t}. Moreover, beyond the temperature dependence of material parameters (both thermal and mechanical), an explicit coupling between the thermal and mechanical field could occur \cite{DreStr93a,KovVan15a}. Consequently, the description of ballistic heat conduction requires a model with multi-level couplings, either on the level of the constitutive equation or the balance equations, including the state dependence of parameters. That multi-level way of thinking is also characteristic in the GENERIC approach \cite{Grmela2018b, PavEta18b}.

Later on, the famous result of Guyer and Krumhansl - the so-called window condition - significantly helped the experimental research to find the optimal frequency for the excitation and make visible the wave phenomenon of heat conduction in solids too \cite{GK64, GK66}. Unfortunately, such window condition for ballistic propagation does not exist. As McNelly's Ph.D. thesis reflects \cite{McN74t}, it is difficult to observe the ballistic modes experimentally. The successful observations required extremely pure crystals \cite{Wal63}, mostly made from NaF. Moreover, the appearance of the ballistic signal is very sensitive for the temperature, probably due to the state dependence of material properties, including the thermal expansion coefficient. Sadly, there is no accurate data on the thermal and mechanical properties. Furthermore, its accurate quantitative reproduction using a computer simulation is almost impossible since temperature scales are missing from the experimental data in many cases, beyond knowing the proper material parameters \cite{KovVan16, KovVan18}.

Despite these difficulties, a few authors have achieved results in modeling ballistic propagation, together with second sound. Ma \cite{Ma13a1, Ma13a2} extended the complex viscosity model of Rogers \cite{Rog71a, Rog72a} and Landau \cite{LandauVIeng}, first including both the longitudinal and transversal ballistic modes, with moderate success. Dreyer and Struchtrup \cite{DreStr93a} applied the RET framework \cite{MulRug98b, RugSug15b}, a phonon hydrodynamical model in which three momentums are considered. Despite the improper propagation speeds appearing in their simulations, they provided the first characteristically acceptable reproductions. Later, Kov\'acs and V\'an \cite{KovVan18}, using the internal variable framework \cite{BerVan17b, JozsKov20b}, compatible with the RET model \cite{KovEta20a}, quantitatively reproduced two series of experiments with predicting the temperature dependence of relaxation time.

%--------------------------------------------------------------------
\subsubsection{Room temperature experiments}
Analogously to the macro-scale low-temperature situation, ballistic and second sound effects could appear both in rarefied and nano-systems under room temperature conditions \cite{Kov18rg}. That is more visible through the eye of kinetic theory, in which the Knudsen number characterizes the `rareness' of the system. For instance, phonon hydrodynamics is applicable for processes with `high' Knudsen number (usually, high means $>0.01$) \cite{Struc05}. In such a situation, second sound and ballistic propagations become experimentally visible. A rarefied phonon gas model is utilized in the RET approach to understand these phenomena. This problem is analogous to gases of real molecules at low-pressure states in which the mean free path becomes large enough to reach the limit for observation and has a significant influence on the transport process.

\paragraph{Rarefied gases}
Based on the computer simulations of ballistic conduction, it has become apparent that the proper coupling between the heat flux and thermal pressure stands as an indispensable requirement to include the ballistic effects. In that particular case of rarefied gases, the situation remains the same, except that the pressure is now the complete mechanical pressure. Consequently, one needs the generalization of the Navier-Stokes-Fourier system, in which it is worth separating the deviatoric and spherical parts of the pressure tensor \cite{Meix43a, AriEtal12c, Arietal15, Arietal13, KovEta20a} since they represent different couplings due to their different tensorial order. Consequently, the deviatoric and spherical parts have a separate time evolution equation with different relaxation times. This is a key point here since the compressibility attributes are crucial, hence the spherical part could possess a significantly different time scale than the deviatoric one \cite{Arietal13, Kov18rg, Struc12, StrTah11, Struc04, StrucTorr08}.

The ballistic contribution is visible at very low pressures, under around $100$ Pa, and observed as a change in the speed of sound with respect to the mass density variation \cite{Rhod46, Gre56, MeySess57, SluiEtal64, SluiEtal65}. Therefore, the mass density dependence of the material parameters and other coefficients (the thermal conductivity, shear and bulk viscosities, relaxation times, and all the coupling parameters) must be implemented in a particular way. That specificness originates from the kinetic theory: one of the early results in kinetic theory argues about the constant-mass density independent-behavior of viscosity and thermal conductivity, which necessarily leads to non-zero viscosity at zero pressure. This is not proved experimentally but based on an extrapolation to the zero density \cite{GrackiEtal69a}. Different experiments proved its opposite \cite{IttPae40}. That kind of 'contradiction' resulted in the differentiation between the 'physical' (i.e., theoretical) and 'effective' (measurable) viscosities \cite{Kov18rg}. Later, in kinetic theory, Knudsen number-based corrections are appeared for viscosity, based on measurements, both in the dense and rarefied domains \cite{BesKar99, RooDar09}. Moreover, considering ideal gas equations of state, one obtains a particular scaling property of the model in the dispersion relations: the propagation speed depends on the frequency/pressure ($\omega/p$) ratio. However, that scaling property is immediately violated if any of the material parameters is a function of mass density or other equations of state are applied, even in the classical case of the Navier-Stokes-Fourier system.

Consequently, the complete success of the experimental evaluation depends on the proper mass density dependence in the model. Using a continuum theory, we note the freedom to choose the form of each coefficient. Basically, it offers infinitely many possibilities that can be narrowed down with the kinetic theory, for instance. However, utilizing an internal variable framework, it turned out that $\omega/p$ scaling is not necessary to evaluate the rarefied gas experiments, only if all the frequency and pressure data are known, which is the most likely situation since these are controlled parameters together with the temperature \cite{Kov18rg}. Thus the continuum approach offers multiple possibilities to evaluate a measurement, depending on the mass density dependence. This also holds for the EIT framework \cite{CarMorr72a, CarMorr72b, KovJouRog19}.

\paragraph{Nano-scaled experiments}
Due to the microscopic spatial scale, the experimental aspects change. The interpretation of material parameters and state variables becomes more difficult, also the measurements themselves. For instance, temperature loses its primary role. Instead, the thermal and electrical resistances are more accessible to measure \cite{ChanEtal08}. Returning to the heat pulse experiments, it is also possible to conduct such measurement with thin films \cite{CepEtal15}, however, on much shorter time scales. While the heat pulse lasts around $1$ $\mu$s for low-temperature experiments, in that case, it is decreased to $100-200$ fs. The usual thickness of a nanofilm is $200-2000$ \r{A}, and its temperature change induces the transient behavior of surface reflectivity on the time scale of picoseconds \cite{Broetal87}. Recently, the group of Siemens, Hoogeboom-Pot, and Lee et al. \cite{SiemEtal10, HoogEtal15, LeeEtal15} obtained convincing results about observing a ballistic propagation at room temperature. The next difficulty emerged when the size-dependence of material parameters is observed for nanomaterials (e.g., for films and tubes) \cite{Wang11nonFou, Klemens01, CahillEtal02, CahEtal03, KimEtal07, RawEtal09, AlvJou07KN, ChoiEtal06, YangZhangLi10, ChenEtal08, FujiEtal05, Cauguo07}. In particular, the so-called superlattices possess a specific behavior: their thermal conductivity non-monotonously depends on their period thickness \cite{SahaEtal11, SahaEtal16, VazVanKov20}. This is probably caused by the parallel heat conduction channels, each having different characteristic spatial scales. Its quantitative modeling is possible but is still an open question \cite{VazVanKov20, LebEta11a, LebEtal11}.

\paragraph{Heterogeneous materials}
Almost every real - not ideal - material is heterogeneous in some sense, e.g., due to the presence of porosity, material composition, and artificially created inclusions. Although there are a few famous experiments in the literature, such as the ones performed by Mitra et al. \cite{MitEta95}, and Kaminski \cite{Kam90}, those data could not be reproduced by any others and thus widely criticized \cite{Ant05meat, GraPet99, HerBec00, HerBec00b, BriZha09}. Usually, the deviation from Fourier's law is tried to be found in a waveform, similarly to second sound, and modeled with the Maxwell-Cattaneo-Vernotte equation. Unfortunately, none of the experiments show temperature waves, and in most cases, the MCV equation failed to explain the results. However, there are also exceptions: despite the lack of heat waves, the MCV equation could be useful \cite{Jiang03, Banetal05}, but this is not the general case. When the MCV equation fails, the (DPL) models are generally considered. The DPL concept lies on a Taylor series expansion of a constitutive equation, with violating basic physical principles due to various reasons \cite{FabEtal16, Ruk14a, Ruk17a, RukSam13a, FabLaz14a, FabFra14a, Quin07, DreEta09a, KovVan18a1}. Despite its popularity, especially in the biological literature, the DPL model cannot be the following standard equation in the engineering practice after Fourier's law due to its numerous shortcomings. 

The usual way to model heterogeneous materials is to implement every particular information about the material structure as much as possible. For instance, in biological situations, the spatial distribution and placement of artery-vein pairs are built-in, together with the requirement of the exact knowledge about the velocity field of blood flow \cite{ChenHolm80, WJL84, Jiji09b, WJ85, Wulff74}. In other cases, such as foams, a statistical approach is more popular, also for the same motivation. However, such detailed modeling of experimental results is not possible due to our limited knowledge, solely in exceptional situations. There is a dedicated research direction to derive and apply an effective thermal model on various material types\footnote{Find here: \url{irrev.energia.bme.hu}}. An experimental campaign is performed to investigate various materials with the heat pulse measurement technique, such as several rocks, metal foams, and 3D printed samples \cite{Botetal16, Vanetal17, FulEtal18e, FehEtal21, FehKov21}. Rocks are outstanding because they consist of different porosity, micro-crack distribution, and their composition is also varying. In these experiments, a particular non-Fourier phenomenon can be observed, different from the others; this is called over-diffusive propagation, see Figures~\ref{Fig:res1} and~\ref{Fig:res2} for details.

\begin{figure}
	\centering
	\hspace{-1.4cm}
	\includegraphics[width=\columnwidth,height=0.55\columnwidth]{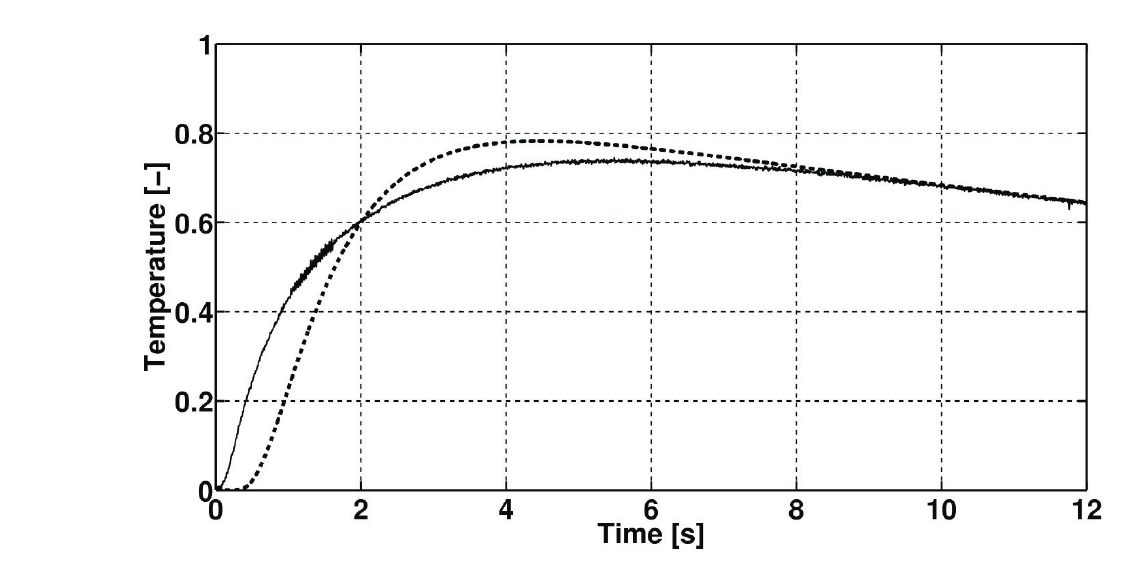}
	\caption{Rear side temperature history from a heat pulse experiment on a metal foam sample, and its evaluation using the Fourier equation \cite{myphd2017}.}
	\label{Fig:res1}
\end{figure}

\begin{figure}
	\hspace{-0.6cm}
	\centering
	\includegraphics[width=0.92\columnwidth,height=0.5\columnwidth]{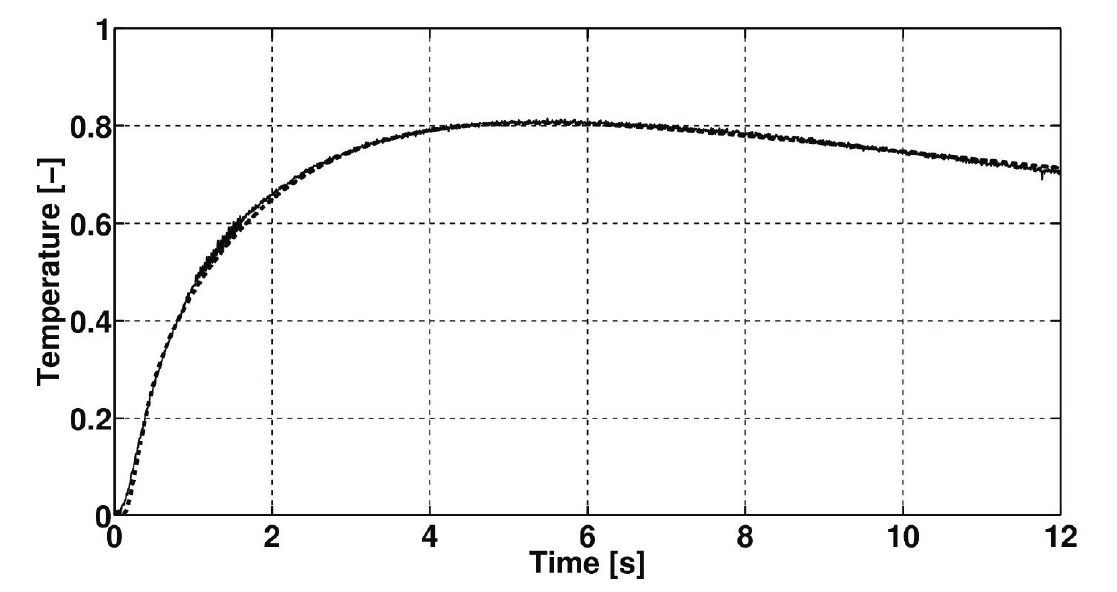}
	\caption {Rear side temperature history from a heat pulse experiment results on a metal foam sample, and its evaluation using the Guyer-Krumhansl equation \cite{myphd2017}.}
	\label{Fig:res2}
\end{figure}

Apparently, this is not a wave-like phenomenon but a diffusive one with multiple conduction channels. That sort of deviation characteristically occurs in any of these heterogeneous materials. This series of measurements is further motivated by the Guyer-Krumhansl (GK) equation, in which the Fourier equation appears together with its time derivative. {That equation is obtained from~\eqref{genhq} if $l_1=0$, i.e., it is part of the ballistic heat conduction equation. Expressing the partial differential equation for the temperature in one spatial dimension with the help of~\eqref{ebal} it has the form
	\begin{align}
		\tau \partial_{tt} T + \partial_t T = \alpha \partial_{xx} T + l^2 \partial_{txx} T \label{GKtemp}
	\end{align}
	where $\tau$ is the relaxation time, $\alpha= \lambda/(\rho c)$ and $l^2$ being a dissipation parameter.} When $l^2/\tau=\alpha$, the solution of the Fourier equation is recovered, this is called Fourier resonance condition \cite{VanKovFul15}, and no deviation takes place. However, when the ratio of $l^2/\tau$ differs from the thermal diffusivity $\alpha$, then the two conduction channels (and their characteristic time scale) differs from each other, the Fourier resonance ceases, and the deviation becomes observable. Furthermore, this also reflects the existence of different time and spatial scales of the constituents. That model is successfully applied for rocks, foams, and biological materials, too. 

Analogously, moisture diffusion problems are similar, both on mathematical and physical levels \cite{Brinkman49, DurBra87}. For example, Wong et al. \cite{WongEtal99} managed to experimentally observe a surprisingly similar deviation than in the case of over-diffusive heat conduction. This is repeated for several compounds, and they also concluded that parallel diffusion processes are the source for such a phenomenon, occurring in heterogeneous materials \cite{KeeEtal05}. For a more comprehensive review of experiments, we refer to \cite{JozsKov20b, Mail2019}.

%-------------------------------------
\subsection{Notes on the solution methods}
We have seen several heat conduction models with different concepts in the background in the previous sections. In order to make them applicable for practical engineering problems or even evaluate an experiment, one needs to utilize a reliable solution method that can reflect their physical content accurately. However, that task carries a lot of physical and mathematical aspects, such as the treatment of initial and boundary conditions. They originate in the structure of constitutive equations. While Fourier's law provides equality between the heat flux and the temperature gradient, the generalized constitutive equations are partial differential equations themselves and may contain further spatial derivatives as well. It affects the definition of boundary conditions, e.g., it is no longer possible to define the flux-type boundary condition using the temperature gradient alone. This is a key point in solving generalized models that must be answered \cite{Kov18gk, BallEtal20, FehKov21}, and cannot be avoided when an infinite spatial domain is considered \cite{Zhu16a, ZhuSri17, Zhu16b}.

{The solution for that difficulty appears differently in analytical and numerical methods. When one chooses to solve the system of partial differential equations analytically, there are multiple possibilities. First, one can decide which state variable will be the primary one, i.e., in generalized heat equations, the temperature and the heat flux are the possible choices. These are called temperature and heat flux representations of the same model. For instance, Eq.~\eqref{GKtemp} is a temperature representation of the GK model. Consequently, in this case, the other variable is eliminated. For example, solving the GK equation for a heat pulse boundary condition, i.e., with time-dependent heat flux on the boundary, it is worth using the heat flux as a primary field variable. After having the solution of the heat flux, the temperature history can be recovered using the balance equation \cite{Kov18gk}. That methodology works for higher-order systems, too \cite{BallEtal20}.

The other way is that no primary field variable is chosen, and thus no variable is eliminated. In that form, the Galerkin solution method is preferred. Similarly to the variable separation methods, the solution is represented by a complete orthonormal set of functions. Sine and cosine functions are a suitable choice. The nonzero boundary conditions can be separated with an auxiliary function, and the homogeneous boundary conditions are satisfied. Moreover, in that way, it becomes possible to implement Robin-type boundary conditions, too \cite{FehKov21}.

The numerical treatment of the boundaries is also tricky due to the same reason. Its advantage is that various types of boundary conditions and geometries are possible to handle without significant inconvenience, but the implementation of boundary conditions is different. For instance, all field quantities are placed on the nodes in conventional finite element methods. Therefore, such discretization requires the parallel and compatible definition of boundary conditions for all field variables. In other words, when a time-dependent heat flux is prescribed on the boundary, it is unknown how to calculate the temperature at the same node in parallel with the heat flux. At this moment, this is an open question for nonlocal models, and thus such allocation of field quantities is not advantageous. Since the phase lag models have only memory-type generalizations, they are more easily fit to the framework of finite element methods \cite{ManMan99, XuLi03, BarSte05a, BarSte08, RahEtal12, VishEtal11, BargFav14}.

Regarding the initial conditions, the methodology also differs from the usual one, independently of the level of generalization. In the case of non-Fourier equations, the initial temperature history is not enough, one other initial condition is necessary. It could be the initial heat flux distribution or the initial time derivative of the temperature. Only the latter can be valid when using the temperature representation of a heat equation. In the case of a nonequilibrium initial condition, the initial time derivative must be compatible with the constitutive equation.}

\begin{figure}
	\label{GK1Ddisc}
	\centering
	\includegraphics[width=0.8\columnwidth]{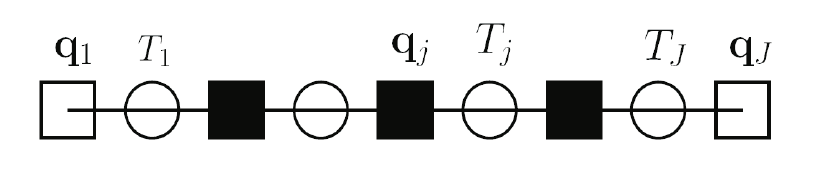}
	\caption{The 1D discretization of the Guyer-Krumhansl equation.}
\end{figure}

In \cite{RietEtal18,FulEtal20}, the modeling capabilities of COMSOL v5.3a is tested on generalized heat equations and compared to an analytically validated numerical method \cite{RietEtal18}. It turned out that nonlocal terms (e.g., the Laplacian of the heat flux in the GK equation) make the appropriate physical content unattainable with this method, independently of the time-stepping and asking a notable computational resource. This attribute is entirely on spatial discretization. That problem can be addressed using a staggered field in space that makes the finite differences similar to the finite volume techniques, except that the balances remain in their strong form. Hence the field variable for which the boundary condition is defined remains on the boundary, but all the others are shifted inside the domain by a half step. These represent the volume average of the small cell, see Figure~\ref{GK1Ddisc} for the concept details. We note that this method demands a thermodynamically compatible structure of constitutive and balance equations, constrained by the second law of thermodynamics. Furthermore, when the physical situation requires boundary conditions for more than one field variable, as an example for a thermo-mechanical problem, it is still doable without any further restrictions if the structure of the equations allows that step. This is independent of the coordinate system and realizable, e.g., in cylindrical coordinates as well.

Finally, we mention the importance of time-stepping algorithms. Together with the spatial discretization, the numerical scheme can produce artificial errors like dissipation and dispersion, independent of their accuracy \cite{NumRec07b,JozsKov20b,FulEtal20}. For diffusive problems, these are hardly visible, and only a thorough analysis can show their presence. However, for wave propagation - especially for elastic waves in ballistic heat conduction or in purely mechanical problems; they have outstanding importance in obtaining an efficient and reliable solution. Here, the difficulty arises from the physical content: elastic waves are non-dissipative, thus the total energy of the wave must be preserved. Consequently, the numerical scheme must satisfy that requirement and must be free from dissipative and dispersive errors. This is achievable even with a first-order accurate time-stepping method, called semi-implicit Euler, the simplest realization of symplectic integrators \cite{Romero10I, Romero10II, PortEtal17, ShanOtt20}. Its realization is as easy as any other finite difference method, but the order of updating the field variables becomes essential. Despite the low accuracy, even this method can be satisfactory. Moreover, conserving the total energy is also advantageous for dissipative systems, therefore the solution also preserves this strong physical property \cite{FulEtal20}.
	
We have seen that theoretical requirements not only helpful but essential to solve the equations and this way to evaluate experiments. Neglecting or disregarding these aspects may lead to instabilities, unphyiscal solutions and entirely prevents the reliable application of the modeling equation.

%-----------------------------------------------------------------------------
\section{Bioheat lagging models}
\label{bio}

The DPL model with the two defined phase lag parameters is found to be an eligible method for heat conduction modeling in the linear regime \cite{Tzo95a,Tzo97b}. The heat flux relaxation time $\tau_q$ and the temperature gradient phase lag $\tau_T$ defined in the DPL model impact the thermal responses. In recent years, much attention has been devoted to non-Fourier bioheat transport, where the DPL model is combined with the existent bioheat equation to model the bioheat transfer processes accurately. This is in contrast to the enumerable studies in other fields of heat conduction such as thermal transport in transistors \cite{Ghazanfarian2009,Shomali2012,Samian2013,Samian2014,Moghaddam2014,Shomali2014,Shomali20152,Shomali2016,Shomali2017,Shomali2018}. { It is remarkable that in almost all bioheat calculations, only the differential version of the DPL constitutive equation is applied. Although the DPL model has the same T-representation as the Guyer-Krumhansl equation, their constitutive parts differ, therefore, they also differ in their physical meaning, which has numerous consequences on their interpretation and solution method, too.
In the following, we do not change the original terminology of the publications.}

Such an increase in bioheat studies can be attributed to two related reasons. The first one is very much associated with the advancing technology in medicine. This improvement makes the required experimental data more accessible, resulting in the acquisition of reliable methods for predicting thermal behavior in living tissues. The second reason is that advanced modeling can greatly help biomedical treatments and the development of new techniques. From a medical point of view, a thorough comprehension of thermal responses of biological tissues is required to ensure the patients' safety during hyperthermia and cryotherapy. The study of skin bio-thermomechanics is also essential for military and space operations to provide astronauts and army personnel with complex clothes for thermal protection. { The above examples show, that  bioheat studies are prominent and non-negligible in many aspects of human life. Here we review the most recent literature regarding this consequential topic. }In particular, the most challenging topics in bioheat heat transport considering the DPL heat conduction model within the last five years are presented.

%-------------------------------------------------------------------
\subsection{Thermal treatment}
Many studies have been devoted to the investigation of the momentous topic of hyperthermia treatment of biological tissues. To be more precise, nowadays, the thermal therapy of different unhealthful organs of the human body has been the topic of numerous inquiries. In a research project, the treated forearm has been studied, solving the Pennes Bioheat Transfer Equation (PBHTE) \cite{Shirkavand2019}. The PBHTE model is among the first bioheat thermal models, considering several biological effects as source terms in the energy balance. For instance, metabolism and blood perfusion are the most frequent terms.

Generally, by decreasing the body metabolism, increasing the blood perfusion rate in tissue, and applying a fluctuating heat flux, instead of uniform heat flux on the surface of forearm skin, it is feasible to optimize the thermal treating of the damaged tissue without causing lesion such as burn injuries to the other healthy parts. This study has not contemplated the non-Fourier heat transfer behavior in the living tissues, resulting in under/over-estimating the calculated parameters. An important step forward is adding the phase lagging phenomenon to the PBHTE model. Work on modeling the heat transport in complex organs with the DPL model has been lately done \cite{Ciesielski2020}.

In more detail, the thermal processes in the polyp-colon system during the electrosurgical polypectomy are studied. This medical surgery technique is usually carried out simultaneously as the colonoscopy is used to remove the abnormal growths from the colon (the large intestine). During this electrosurgical polypectomy procedure, a polypectomy snare is fastened around the stalk of the polyp. Then, the electric current will be applied to the snare for a few seconds. This causes the polyp stalk to be detached from the colon wall. The PBHTE differential equations describing the processes during the polypectomy have been solved using the control volume method. The simulation results help the endoscopists by letting them know the optimal time and parameters of the electric current flow depending on the geometry and size of the polyp. More fully, the predicted rate of the thermal damage of the tissue obtained by the Arrhenius damage integral, the electric field distribution profile, and the thermal processes occurring in the polyp-colon domain, can be utilized to optimize the endoscopic instrument parameters for the clinical applications. As mentioned before, the bioheat transfer problems in the living biological tissues should be simulated using the bioheat models based on the non-Fourier heat conduction, i.e., the GK or the DPL model.

In the mentioned work, these procedures have not been considered. The authors indicate that the lack of experimental data of the phase lag times for the colon and polyp tissues are the main difficulties, but they confirm that the non-Fourier models will be taken into account in their future works. Here, we should mention the experimental study by Liu \emph{et al.} in which the phase lag times for the living tissues are predicted \cite{CLi2018}. Various DPL applications for thermal therapy of ill tissues are presented in the following.

%---------------------------------------------------------------------
\subsubsection{Modeling of radiofrequency (RF)/microwave(MW) ablation}
Latterly, an initiative more accurate method for modeling the RF/MW ablation remedy of the cancerous tissues has been proposed by Singh and Melnik \cite{Singh2019,Singh20202}. A coupled thermo-electro-mechanical model has been developed while considering both tissue shrinkage and expansion. Also, the coupled model considers the non-Fourier effects by including the single-phase-lag (SPL) and dual-phase-lag (DPL) models of bioheat transfer during the high-temperature thermal ablation by MWA and RFA. The authors have reported that neglecting the mechanical coupling during the modeling of MWA results in an underestimated temperature distribution. It also emphasizes the significant contribution of the mechanical terms in the internal energy. Furthermore, there is a notable deviation between the predicted damage volume acquired from the coupled thermo-electro-mechanical model of MWA considering Fourier and non-Fourier models. 

Additionally, the effect of non-Fourier-based coupled thermo-electro-mechanical coupling is reported to be less pronounced in RFA as compared to the MWA. However, using the newly proposed model could lead to more accurate predictions of the temperature distribution and the damage volume during the RFA. In addition, thermal ablation as a non-aggressive method is often used to destroy the defective tissues in the heart. In both thermal RF ablation and cryo-ablation, this technique is utilized to treat heart arrhythmia. Newly, a two-dimensional coupled thermo-electro-mechanical model has been developed and employed to treat the heat process during the thermal catheter ablation. The model incorporating both tissue contraction and expansion during the MWA and RFA procedures considers DPL-BHTE along with Helmholtz harmonic wave equation, the modified stress-strain, and the thermo-elastic wave equations \cite{Singh2019}.

The agreement with experiments can be improved by taking into account the temperature dependence of transport coefficients (e.g., electrical and thermal conductivities). It increases the importance of a physically and mathematically consistent thermodynamic theory. Otherwise, the results could be misleading and cannot be applied to further problems. The obtained temperature profile is found to be smaller than that of the Fourier prediction. This finding is in contrast to what was obtained in \cite{Kumar2016ll,Zhang2017}. 

%--------------------------------------------------------------------

\subsubsection{Magnetic hyperthermia}
The one-dimensional dual-phase-lag bioheat transfer model for the bilayer living tissues during the magnetic fluid hyperthermia treatment is modeled using the finite element Legendre wavelet Galerkin method (FELWGM) in \cite{Kumar2016ll}. This model can be used for an effective MFH treatment of bilayer tissues. The FELWGM converts the contemplated problem into a system of algebraic equations. The obtained results present no notable difference for the Fourier or non-Fourier boundary conditions.

The effect of time delays only appears in the tumor region. In detail, the temperature increase is characterized by $\tau_q$ and decreases with the augmentation of $\tau_T$ in the tumor zone. This study suggests FCC FePt magnetic nanoparticle (MNP) as the most effective MNP used for the thermal treatment. The effect of various magnetic heat source parameters such as magnetic induction, frequency, the diameter of magnetic nanoparticles, the volume fraction of magnetic nanoparticles, and ligand layer thickness has been investigated. The physical property of these parameters has been described in detail during the magnetic fluid hyperthermia (MFH) treatment, and also the clinical application of MFH in oncology is discussed. 

A nonlocal dual-phase-lag (NL-DPL) model, which provides the effects of thermomass and size-dependent thermophysical properties at nanoscale, is also developed for MFH treatment \cite{Kumar20164,RKumar20192}. The effect of size-dependent characteristics, the Dirichlet, Neumann, and Robin boundary conditions, and the phase lag parameters in the tumor and normal zones of the bilayered structure under the MFH therapy are discussed. The finite difference scheme and Legendre wavelet Galerkin technique, which modifies the problem into a system of algebraic equations, are utilized. It is obtained that as the characteristic length increases, the temperature also grows. In other words, this work confirms that the NL-DPL model is more realistic than the usual DPL model at nanoscale. The nonlocal parameters are found to affect the first layer consisting of the cancerous cells, which are exposed to the injected magnetic nanoparticles. The NL-DPL model predicts that the treatment time, an essential issue for thermal therapy is less than that of the DPL model. Although this study lacks verification, it can be practically beneficial due to the annunciation of the thermal therapy time reduction, an important parameter for patients' comfort.

Recently, K.C. Liu and his colleagues have also studied magnetic hyperthermia under the dual-phase-lag platform \cite{KCLiu2018,KCLiu2019,KCLiu2020}. Moreover, the generalized dual-phase-lag model of bioheat transfer has been developed to investigate the heat transfer inside the living tissues induced by a Gaussian spatial source. Furthermore, a hybrid numerical scheme is developed, combining Laplace transform with hyperbolic shape function technique in order to ease the implementation of complex geometries and interfaces with singular points. Regarding the blood temperature, a criterion is developed that limits the power dissipation in magnetic hyperthermia for safe treatment.
Also, these studies reveal that due to the Gaussian distribution of magnetic particles, it is not easy to obtain the temperature distribution within the tumor.

It has been illustrated that a longer exposure time is required for complete ablation of the tumor, even the tumor temperature has exceeded 42 $^\circ$C. Briefly, the mentioned studies solve the generalized dual-phase-lag bioheat transfer model and point out that the heating time and power density are critical to prevent thermal damage and must be controlled in hyperthermia. Although the scrutinies by Li and his coauthors answer the challenges for solving the bioheat transfer model for concentric bilayered solid spheres, developing full models considering the existent fluid dynamics in the organs can be taken as the next step. At last, it should be noted that as there exist significant challenges in finding the reliable temperature profiles inside the living tissues during the MNP hyperthermia treatment, the development of reliable and accurate numerical models is indispensable \cite{Raouf2020}. On the other side, the temperature profile in the living tissues during the MNP hyperthermia therapy is determined using the nonlinear form of the DPL model \cite{HMYoussef2019}. The concentration of the injected magnetic particles is experimentally found to be a Gaussian distribution, and the governing PDE is solved in the Laplace transform domain. It is presented that the tumor, the tissue, and the thermal damage quantity are effectively affected by the nonlinear and linear effects of the phase-lag times.

%--------------------------------------------------------------------
\subsubsection{Modeling the treatment through Focused Ultrasound (FU)}
Thermal treatment of the benign thyroid tumor with FU has been recently numerically investigated \cite{Namakshenas2019}. In greater detail, the ultrasound irradiation with different powers of 3 W, 5 W, and 7 W and the frequency of 3 MHz has been applied to the multilayer model of the neck consisting of the internal organs, extending from the skin toward the thyroid gland, and constructed from the CT-scan images. It is stated that the considered number of layers has a significant effect on the calculated parameters.

For instance, the obtained maximum acoustic pressure in the multilayer model is 1.18 times more than that of the common two-layer model, where the two layers of water and tumoral tissue are considered only. The thermal wave, the DPL, and traditional Pennes bioheat transfer models are solved to study the temperature profile. For the power 3W, the temperature distribution obtained by the non-Fourier thermal models presents the maximum temperature with time delays of 11.32, 5.66, and 2.86 s that are 20.51$\%$ 14.1$\%$ and 8.65$\%$ lower than that of the Fourier model. As the power of the transducer increases, the deviation from the Fourier results also increases. Also, the effect of the phase lags on the area near the necrotic tumoral zone is studied. It is figured out that the region with irreversible thermal damage shrinks for the 3~W power and 5.66 s time delay and no necrosis of the thyroid nodule occurs. Besides, in another recent study, the proposed model is improved by considering the effect of the interfacial convective heat transfer between the blood vessels and the extravascular matrix \cite{Namakshenas2020}. Also, very recently, High Intensity Focused Ultrasound (HIFU) as an appropriate treatment for thermal ablation has been proposed in \cite{Singh2020}. An effective HIFU thermotherapy for precisely predicting biological tissue temperature profile has been announced. Moreover, in-vitro experiments were performed to validate the numerical results.
%--------------------------------------------------------------------------------

\subsubsection{Laser therapy}
The laser interstitial thermal therapy (LITT), a novel technique for the treatment of primary and metastatic tumors, has been simulated using the DPL model \cite{Mohajer2016}. Similar to the other thermal treatments, the efficiency of the method depends on the success of the temperature control, which relies on the living tissues' thermal properties. In particular, the DPL equation has been investigated numerically. Furthermore, the geometrical structures affect the thermal response as well as the overshooting phenomenon in biological tissues. It should be noted that the modeling of the more realistic cases of study, including the consideration of the blood vessel in the tissue and also the non-uniform perfusion are desirable for obtaining accurate prediction of the LITT result.

The human tooth composed of enamel, dentin, and pulp with unstructured shape, uneven boundaries, and realistic thicknesses under the effect of laser irradiation has been modeled using the DPL heat conduction model \cite{SFalahatkar2017}. It has been concluded that the heat flux phase lag significantly affects the temperature profile at early stages, while the temperature gradient phase lag is more important at later stages. The outputs are validated with the experimental results by taking $\tau_{q}=16 ms$ and $\tau_{T}=2 ms$. Besides, the authors have published a work in 2018 reporting the simulation of laser irradiated tooth with the three-phase-lag bioheat transfer model. The phase lags are found to be $\tau_{q}=16 ms$, $\tau_{T}=6 ms$, and $\tau_{\nu}=2 ms$ \cite{Falahatkar2018}. In the first work mentioned above, the DPL model with the determined phase lags is introduced to be the suitable model for the description of the non-Fourier heat transfer in teeth. Therefore, the debate arises: what is the necessity of the second research \cite{Falahatkar2018}.

In other words, it is not yet established which non-Fourier model, TPL or DPL, is better in the prediction of the experimental results. Further, these studies use different $\tau_{T}$s for the two models of TPL and the DPL, which is also debatable. Dealing with thermal therapy of the damaged tissues, the cancerous skin tissue ablation has been numerically investigated \cite{Nobrega2017}. Laser ablation as an efficient method of thermal treatment for tumorous tissue increases the temperature of the tumor cells. Consequently, irreversible burn and deformation result in protein denaturation and cell membrane dissolution, which themselves destroy the cancerous cells. Three models (Pennes, thermal wave, and DPL) have been used for the investigation of two different laser irradiation approaches of the tissue, which highly depend on the absorbent or scatter feature of the skin.

These three models, solved with a finite element method, are reported to present a significantly different temperature profile. More precisely, it is stated that heat flux and temperature gradient phase lags, and thermal conductivity of the tumor affect the results substantially. On the other side, the specific heat and the blood perfusion rate have a negligible impact on the thermal damage. Although the verification is presented for the simplest available case, an experimental validation for the specific situations of tumor ablation is still missing. This expresses the severe desideratum for reliable, optimal in-vivo experiments trying to overcome the limitations such as the injury caused. A precise study modeling the laser-mediated therapy for port-wine stain (PWS), which is a vascular skin disease caused by congenital vascular deformity, is performed in \cite{Zhang2017}. The laser energy is absorbed mainly by the blood's hemoglobin to form clots, destroying the abnormal capillaries. The optical transmission and energy deposition are calculated via Monte Carlo-based method, and the non-Fourier heat conduction is treated solving the three-dimensional DPL model with the temperature jump boundary condition \cite{Ghazanfarian2009,Shomali2012} model via using a three-level finite difference method.

It is presented that $\tau_q$ and $\tau_T$ result in lower temperature distribution than that of the estimations using Fourier's law. This dependency is more pronounced when the delay times are more significant. Also, the effect of $\tau_q$ is more notable while the $\tau_T$ equalizes the temperature gradients at the interfaces \cite{Kumar2016ll}. Such inconsistency can be attributed to the different dimensionality in these two studies and the different types of the studied therapies.

%--------------------------------------------------------------------
%--------------------------------------------------------------------
\paragraph{Applications with gold nanoparticles}
It is confirmed that the thermal investigation of the biological tissue phantom showing an optical inhomogeneity and embedding with the gold nanoshells, under the effect of the pico-second laser irradiation, has been numerically performed in \cite{Phadnis2016}. The absorbing optical inhomogeneity demonstrates the presence of the malignant cells. The gold nanoshells have been uniformly dispersed over the body of the inhomogeneity to enhance the absorption properties while maintaining the optical properties of the background at the reference values. The generalized form of the Radiative Transfer Equation (RTE) has been applied to model the light-tissue interaction using the discrete ordinate method (DOM). Then, the obtained intensity distribution is utilized to solve the generalized form of the non-Fourier heat conduction model, i.e., the DPL bioheat transfer model. It is reported that adding nanoshells to the optically inhomogeneous zone results in notable changes in optical properties leading to the localized increase in the temperature of the embedded inhomogeneity.

Numerical and experimental investigation of the heat transport in the collagen microstructures under the NIR laser irradiation, both for bare mimics and with gold nanostructures infused, has been worked out in \cite{Sahoo2018}. The non-Fourier DPL model incorporated with the Pennes bioheat transfer equation is solved using the COMSOL multiphysics software. The deviations of the maximum temperature calculated from the DPL and the Fourier models from the obtained experimental value in pure collagen are reported to be 1 and 4 K, respectively. These differences in the phantoms embedded with nanoparticles are respectively 0.5 K and 4 K. Non-Fourier calculation is necessary while treating tissues; otherwise, the finding will be unrealistic from clinical perspectives and lead to the damage of the healthy tissues.

In 2018, a study presenting the effect of the targeted gold nanostructure injection on attaining the precise necrosis of the tumor and leaving the minimum damage to the healthy tissue during the laser thermal therapy had been accomplished \cite{Paul2018}. More closely, combined diffusion and convective energy equations were solved by the usage of the COMSOL Multiphysics software. Furthermore, the heat propagation induced by the laser and together with the thermal damage are modeled by the modified Beer-Lambert law and Arrhenius equation, respectively. Related to the present review, a tumor-blood inhomogeneous inner structure is also studied under the basis of the non-Fourier DPL bioheat transfer approach. The comparison between the non-Fourier results and that of the classical Fourier approach establishes the considerable difference of the temperature profile during the initial phases of laser heating and cooling, i.e., non-equilibrium condition. The outcomes converge at larger time intervals.

The work presented by Yin \emph{et al.} \cite{Yin2020}, the gold nanoparticles (GNPs) enhanced laser-induced thermal therapy has been studied theoretically. The bioheat transfer model under laser irradiation has been solved with the aid of the finite element method. Yin et al.~investigated the effect of numerous parameters such as the laser intensity, the anisotropic scattering characteristics of nanoparticles, convective heat transfer coefficient, and the treatment strategy. It is reported that the oblique and vertical laser incident have the same efficacy for the single-dose therapy, where the whole tumor is damaged in a one-time step. While during fractional therapy, the influence of the oblique laser irradiation is more than that of the vertical one.

%--------------------------------------------------------------------
\subparagraph{Comparison with HIFU}
The thermo-mechanical responses of the living tissue under the effect of continuous and pulse mode heating during HIFU and LITT subjected to intravenous injection of GNPs, are studied in \cite{Paul20202}. Like the other treatments, the goal is to reach maximum necrosis of the tumors with the least healthy tissue destruction and the least nociceptive pain. Accurately describing, a three-dimensional multilayered breast tissue including the tumor and also the multilevel artery and vein has been investigated. The tumor was treated with ultrasound focusing on its location during HIFU therapy, while in the LITT therapy, the thin catheter inside the tumor heats the region.

The lagging time model is dealt with concurrently solving the coupled radiative transfer, Helmholtz, momentum, DPL equations and equilibrium equations for optics, acoustics, fluids, temperature and mechanical fields via COMSOL Multiphysics software. It is affirmed that as opposed to the continuous mode of heating, the tissue temperature rise in pulse mode leads to a better-targeted tumor necrotic damage while keeping the encircling healthy tissues untouched. The non-homogeneous impacts of multilevel artery/vein in tissue are stated to be more significant during LITT in comparison to HIFU.

On top of that, the effect of the artery/vein existence is found to be more pronounced over the pulsed mode of heating. Further, the thermoelastic stiffness of the tissue is lower for the pulse heating, causing less nociceptive pain in comparison to the continuous energy agitation. Moreover, the large blood vessels are declared to be momentous in increasing the stiffness of the tissue. In the presence of the GNPs, the optical and acoustic properties of the tissue are reinforced. This enhancement itself results in modification of the thermal and elastic demeanor of the tissue during the therapy. This study also provides experimental work on agar-based tissue phantoms. The validation of the simulation results of the FU heating is performed using the acquired experimental data.

%--------------------------------------------------------------------
\subparagraph{Pain assessment}
Continuing the previous work, recently, the numerical assessment of the induced nociceptive pain during the thermal ablation has been performed by calculating the viscoelastic deformation via modeling the thermomechanical response of the tissue \cite{Paul2020}. Again, the DPL and Pennes bioheat model coupled to the equilibrium equations are solved using the COMSOL Multiphysics to calculate the thermal and mechanical fields in the triple-layered skin structure. Furthermore, an in vitro investigation of the single-layered tissue phantom has been performed to validate the numerical results. That paper further tackles the question of which intravenous (IV) and intratumoral (IT) infusion of the silica-gold nanoparticles leads to the more precise and uniform necrosis of the malignant tissues during laser therapy. The computation suggests that the IT scheme presents more precise tumor ablation and less thermally induced skin deformation in comparison to that of the IV scheme \cite{APaul20201}. Although this work exhibits novelty in investigating the effect of gold nanoshells' presence on more successfully laser irradiation of the abnormal tissues, uncertain values of the relaxation times have been taken into account during the modeling. Using the newly obtained values \cite{CLi2018} will make the next step to be taken more firmly for achieving more accurate results.

%--------------------------------------------------------------------
\subsection{Thermal protective clothing}
A dual-phase lag model of the bioheat transfer equation is used to anticipate the burning time. Although there are many methods to ascertain the skin burn level at the clinical level, only a few methods are appropriate for assessing the protective clothing. Basically, the methods fall into two categories: empirical criterion prediction and theoretical approaches. An empirical criterion uses the accumulated skin surface energy in specified times to find the onset of second-degree skin burn. The latter is a more complex approach, including a heat transfer and a burn model, which designates the exact degree of burn based on the inner skin temperature.

Zhai and Li have reviewed different burn prediction techniques which are used in clothing \cite{Zhai2015}. These methods are founded on experiments performed on living bodies getting in touch with hot water or exposed to radiant sources. Hence, the protective fabrics between the heat source and skin can make the prediction methods inaccurate due to the block of some incident heat and the fabrics' change of the heat form. Accordingly, new technologies should be used to scrutinize the accurate and efficient prediction methods for both cases of burning. On the other hand, the review dealing with experimental and numerical studies of thermal injury of the skin and subdermal tissues has been written \cite{Ye2017}. The first part of the review concerns experimental research containing burn conventions and prevailing imaging techniques. The existing numerical models for the tissue burn and relevant computational simulations are reviewed in the second part. This review concludes that although many studies are devoted to the simulation of the pathology and pathogenicity of the tissue burn, there is limited information concerning the appearance of deformation in the tissue characteristics, including mechanical, thermal, electrical, and optical properties.

%--------------------------------------------------------------------
\subsection{Cryoablation}
Cryosurgery or cryoablation is a non-invasive technique for the treatment of cancerous tissues. A study using the immersed boundary method (IBM) for simulation of the cryofreezing biological liver tissue while adding a heat source in the bioheat transfer equation has been first performed by Ge \emph{et al.} \cite{Ge2015}. The obtained results for the liver tissue with temperature-dependent thermophysical properties have been shown to present good consistency with the available data from the previous numerical and experimental works. Also, a heat source due to the blood flow in the embedded vessel significantly affects the temperature distribution. In detail, while the distance between the cryoprobe and the major blood vessel varies, the ice fronts 0 $^\circ$C and -40 $^\circ$C also change, up to 35$\%$ under 500 s.

The most critical parameter dealing with DPL bioheat modeling is the determination of the phase lags. In this paper, the $\tau_q$ and $\tau_T$ are both taken to be $0.001 s$. This value is different from what is expected recently, and the reason for such a choice has not been cleared out in the manuscript. Further, in 2017, the problem of biological tissue freezing was modeled with the thermal interactions between the cryoprobe tip and soft tissue being described using the DPL model \cite{Mochnacki2017}. The explicit finite difference method is used, and the freezing process modeling is performed based on the newly defined parameter, called substitute thermal capacity. Despite the previously mentioned work \cite{Ge2015}, the present work has provided a detailed discussion on the suitable values of the relaxation time, $\tau_q$, and thermalization time, $\tau_t$. More, this work has interestingly stated that the increase in the relaxation time makes the differences between DPL results and that of the Pennes solutions larger, while such growth for thermalization time offsets the divergences between the DPL and the Pennes solutions.

This can be justified as the large $\tau_q$ presents a significant lag of the heat flux and consequently makes the thermal accumulation that opposes the thermal equilibrium and results in fluctuation. However, a high value of $\tau_{T}$ demonstrates a conspicuous lag of temperature gradient, decreasing the thermal accumulation and accordingly improving the thermal balance. Hence, larger $\tau_q$, takes away the system from the equilibrium situation and makes the obtained result more deviated from the Pennes' solutions. The accuracy of the method has been verified by comparing the obtained results with the one adopted from the Pennes equation and the experimental data. In the following, a two-dimensional DPL model is developed to study the phase change in a heat transfer process during the lung cancer cryotherapy \cite{Kumara2017}. The governing DPL bioheat conduction model is worked out numerically using the enthalpy-based finite difference method. 
 Contemplating the non-ideal behavior of the tissue, heat source terms, metabolism and the blood perfusion, the study inquires about the effects of phase lags on the obtained heat flux and temperature-dependent parameters during the freezing process.

The phase lags are taken to be  $\tau_q=10 s$ and $\tau_T=10 s$, which are close to the experimentally verified phase lags. Remarkable efficacy on the interface positions and the temperature distribution is reported. Furthermore, it is obtained that among the DPL, hyperbolic and parabolic models, the entire tissue freezing time is the least and the largest, respectively, for the parabolic and hyperbolic model while it presents the moderate value for the DPL model \cite{SKumar2018}. On the other hand, a new modified Legendre wavelet Galerkin method describing bioheat transfer during cryosurgery of lung cancer has been developed \cite{MKumar20181}. The lung tissue cooled by a flat probe with its temperature decreasing linearly with time is modeled. The essential novelty of this paper is the new suggested method for modeling the freezing process. Considering the cooling procedure occurred in three stages of cooling up to the liquid temperature 10~$^\circ$C, cooled up to the freezing temperature 80 $^\circ$C, and cooling up to the lethal temperature with the whole region to be divided into the liquid, mushy and solid zones, respectively.

The problem is converted into a boundary value problem in stage one and the moving boundary value problem in stages two and three. Finally, the problem shifted to the generalized system of Sylvester equations and is solved using the Bartels-Stewart algorithm of the generalized inverse. With the value of phase lags to be unknown, the temperature distribution and moving layer thickness in all stages and regions are traced to predict maximum damage to the infected tissue and the minimum detriment to the healthy lung tissues. While the present calculation is performed for the one-dimensional case \cite{MKumar20181}, the auxesis to the two-dimensional one is carried out in more recent work of the same group \cite{MKumar2019}.

%--------------------------------------------------------------------
\subsubsection{Cryopreservation}
Cryopreservation is a procedure preserving the biological tissues while making the less severe failures to the physical and functional properties. Such preservation is achieved by exposing the tissues to very low cryogenic temperatures involving a more significant heat removal rate. The Pennes bioheat model is used for two-dimensional numerical modeling of the cryopreservation operation. In detail, using the Finite Volume Method for discretization while utilizing the Tridiagonal Matrix Algorithm to solve the discretized algebraic equations, the temperature distribution is obtained \cite{Sukumar2019}. Moreover, the solid-liquid interface is tracked with the Enthalpy-Porosity method during the freezing process. In the abovementioned modeling, two cases with different methods of treating the tissue freezing implying from one side and the two sides are studied.

It is found that the freezing rate of tissue is doubled as the cooling process is changed from one-side freezing to the two-side one. More importantly, the lower value of blood perfusion rate is observed to lead to the lower value of the final temperature of the tissue. In other words, the tissue with a high blood perfusion rate will freeze to the lower cooling medium temperature. The metabolic heat generation has been found to take no substantial role in the temperature profile of the healthy tissue. To conclude this section, we emphasize that these studies' essential and useful extension treats the more realistic cases of three-dimensional systems. In other words, future investigations should be devoted to the more reliable and accurate three-dimensional study of bioheat problems.

%--------------------------------------------------------------------
\subsection{ Bio thermo-viscoelastic/mechanical model}
There have been many works on the viscoelastic properties of biological tissues. However, the coupled thermal and mechanical behavior of the biological tissues based on the viscoelastic theory has been investigated recently, and the dual-phase-lag thermo-viscoelastic model is developed \cite{XLi2020}. Further, the transient behavior in tumor and normal tissues have also been studied. It has been reported that although the absolute values of displacement and stress are smaller in comparison to the obtained values from the conventional thermoelastic model, the viscoelastic parameter has little effect on the value of the temperature.

This means that during the hyperthermia treatment for the cases with the same thickness and time parameters, the thermal deformation damage of the thermoelastic model is more severe than that of the thermo-viscoelastic model. In this study, it is considered that the tumor and normal tissues have linear viscoelastic properties. It has been also emphasized that an extension to the nonlinear generalized thermo-viscoelastic model should be performed to reach more realistic results. On the other hand, the thermomechanical interactions in anisotropic soft tissues are studied using a new boundary element algorithm \cite{MAFahmy2018}. The governing equations are based on the DPL bioheat transfer and Biot's theory. Both equations are solved independently. First, the temperature distribution is found by solving the DPLBHT equation with the Boundary Element Method (BEM). Secondly, the displacement components distributions are obtained by resolving the mechanical equation with the convolution quadrature boundary element method (CQBEM). The CQBEM with low CPU and low memory usage is a suitable technique for handling soft tissues with complicated shapes. The subsequent bioheat and mechanical linear equations are solved via a transpose-free quasi-minimal residual (TFQMR) solver that implements a dual-threshold incomplete LU factorization preconditioner to improve the total CPU time. Good consistency between the obtained BEM results and that of the analytical, FDM, and FEM is reported.

This is true while the computational cost, including the CPU-time, memory, and disc space for the BEM, is much lower than the other methods. Further, a theoretical analysis on the thermo-mechanical responses of the human tissue under the effect of a moving laser beam during the thermal ablation is performed \cite{Ma2019}. The skin's thermo-mechanical responses throughout the therapy are studied using a mechanical model that includes an elastic plate established on a viscoelastic foundation. The model is studied utilizing the DPL bioheat conduction model, the Kirchhoff hypothesis, and the Kelvin-Voigt rheological model. The governing equations are solved analytically using the Green function method. It is found that the increment of $\tau_q$ and $\tau_T$ increases and decreases the deflection and stress magnitude, respectively. In addition, the growing laser moving speed is found to reduce the temperature and deformation magnitude while it extends the thermo-mechanical response zone.

Besides, a thermo-mechanical model taking into account the non-Fourier effects in the skin tissues via the DPL model is developed for the EEDs \cite{YYin2020}. This model anticipates the human tissue sensitivities, the temperature, and the stress distributions, in the EED/skin system. The governing equations are solved using the Green function method. In this work, the phase lag parameters needed for the DPL model are obtained from the two-temperature model and are validated by the experiments. The integrated analytical framework, consisting of the thermo-mechanical model for the temperature and stress, Arrhenius burn model for the thermal damage, modified Hodgkin-Huxley model for nociceptor transduction, and the gate control theory for the pain modulation and perception, is defined to quantify skin pain sensation in terms of the noxious stimuli from EEDs. This study can be considered as the pioneering work dealing with pain management, leading to the less irritating real clinical practice. The effect of the heat flux and temperature gradient phase lags on the magnitude of the deflection and stress has been investigated.

Also, the influence of the moving laser beam speed presents that the increment of the velocity will reduce the magnitude of the temperature and the deformation but enlarge the thermo-mechanical response zone. On the other side, the bio thermomechanical behavior in a cancerous layer within the context of Lord-Shulman theory is studied using the thermo-viscoelasticity model considering varying thermal conductivity and rheological properties of the volume \cite{Ezzat2020}. The Laplace transform technique is employed to solve the problem in the transformed domain. It is claimed that the thermal conductivity and volume unwinding parameters affect the circulations notably.

{ The thermomechanical coupling is a research area where a consistent thermodynamical approach is available and can improve the modeling \cite{BerVan17b,JozsKov20b}}.
% {\color{green} while stating that the present work presents no verification or validation, it should be mentioned that the Lord-Shulman and Green-Lindsay are basically wrong in these cases and connection to rheology is not really considered \cite{BerVan17b,JozsKov20b}}.

%--------------------------------------------------------------------
\subsection{Comparison with experiments}
To ensure that we are stepping in the right way, studies, including the verification with the experiments, are mandatory. The Achilles heel in the context of this area of research is the lack of experimental validation. Among the more than hundreds of papers dealing with dual-phase-lag bioheat studies, there are not too many that consider such verification \cite{Phadnis2016, Sahoo2018,APaul20201,YYin2020,Hooshmand20151,Alzahrani2019}. Although Maillet has mentioned a few relatively old articles dealing with validating non-Fourier models for bioheat transfer \cite{Mail2019}, it is worthwhile to review the most recent ones. In a study, a tissue phantom and ex-vivo bovine liver tissues were heated by the focused ultrasound (FU) \cite{CLi2018}. The case is also studied using heat conduction models such as the Pennes equation, bioheat transfer thermal wave model, and the DPL equation. The findings establish that the Pennes equation accurately predicts the initial temperature increase for a homogeneous tissue phantom while the prediction deviates from the measured temperature with increasing FU irradiation time. Further, the experimental response is closer to the temperature calculated by the non-Fourier models for the heterogeneous liver tissues, especially the DPL model.
In other words, the DPL model can predict the temperature repercussions in biological tissues due to the increment of the phase lag, which specifies the micro-structural thermal interactions. This detailed study inaugurates more accurate clinical treatment plans for thermal therapies. In addition, the temperature profile in the skin tissue has been obtained using the highly nonlinear DPL bioheat transfer model during the thermal therapy treatment of the tumor \cite{Kumar2018}. The hybrid numerical procedure solves the proposed case based on the spatial discretization technique and the Runge-Kutta method. The calculated temperature distribution presents a good consistency with the experimental data. Also recently, Saeed and Abbas have studied the transient phenomena in spherical biological tissue under the effect of the laser heat source \cite{TSaeed2020}. The finite element approach uses the quadratic elements to solve the highly nonlinear hyperbolic bioheat second-order differential equations. The obtained numerical results were compared with the experimental data. The collation demonstrates that the developed mathematical model is an efficacious tool to estimate the heat transfer in the spherical biological tissues. { We also note that for a DPL model, the nonlinearities cannot be accounted reliably due to the lack of thermodynamic background. Therefore, while there are promising results, one needs to keep in mind the shortcomings as well.}

%--------------------------------------------------------------------
\subsection{Newly evolved mathematical methods}
Due to the intense need for accurate and precise prediction of the situation during the hyperthermia treatment of the cancerous tissues to achieve justified temperature control, seeking reliable methods is one of the most serious topics for studies, including non-Fourier equations. Hence, many efforts have also been made to find efficient methods for solving systems of partial differential equations. The general boundary element method is used to solve the 3-D DPL equation fulfilling the Dirichlet or the Neumann boundary conditions and the initial value \cite{Majchrzak2015}. The numerical calculation is verified by comparing the results with the analytical solution of the 1D boundary initial value problem, artificially extended to the 3D one. Furthermore, the heat transfer phenomenon occurring in the 3D domain of the heated tissue is investigated. The values of $\tau_q$ and $\tau_T$ are not yet correctly assessed. For example, Vedavarz \emph{et al.} \cite{Vedavarz1994} predicted that the value of $\tau_q$ for biological tissues lies between 10-1000s and 1-100s, respectively, for the cryogenic and room temperature. 

Consequently, the interval boundary element method for transient diffusion problems may be broadened to the presented general boundary element method. Furthermore, an approximate analytical solution of the dual-phase-lag bioheat transfer equation utilizing the finite element Legendre wavelet Galerkin (FEWGM) is developed in \cite{Kumar20151}. FEWGM minimizes the error and also produces a higher degree of accuracy. With selecting the Gaussian distribution as the heat source, this study affirms that the effect of the $\tau_q$ on the heat transfer process is negligible while $\tau_T$ effectively affects the procedure. This finding is questioned in the present research. The FELWGM is also used to model the DPL bioheat transfer model in the presence of the metabolic and modified Gaussian external heat source \cite{Kumar2016ll}. With selecting the suitable values of the modified Gaussian external heat source parameters, the results are found to present good consistency with the exact solution. Thermal damages are as well observed to depend on the boundary condition. In spite of the previous work, the authors have announced the effect of both $\tau_q$ and $\tau_T$ on the temperature profile at the target area during the therapy for this type of heat source. On top of that, a method with less computational cost has been recently suggested to solve a highly complex nonlinear dual-phase-lag (DPL) equation \cite{TKumari2020}.

The boundary value bioheat problem is converted to an initial value problem using the finite difference method. The Runge-Kutta method is then utilized to find the dimensionless temperature profile. The effect of various properties such as blood perfusion rate and thermal relaxation time of heat flux and different boundary conditions is investigated in this work. The topic of nonlinear DPL has also been investigated in \cite{Arefmanesh2020} where the semi-analytical solutions of the nonlinear DPL using the Galerkin weighted residuals method are acquired. The nonlinearity appears due to the temperature-dependent thermal properties, including the blood perfusion rate, heat conductivity, and metabolic heat. It is concluded that the temperature-dependent metabolic heat generation and blood perfusion results in much higher temperature in the tumor zone. Also, the temperature-dependent thermal conductivity is found to cause the decrement of tumor temperature. Despite the existence of the clinical records, this research only presents a synopsis verification with one analytical result.

%--------------------------------------------------------------------
\subsubsection{Generalized DPL model (GDPL)}
An analytical investigation of the 1D nonequilibrium heat transfer in biological tissues during laser irradiation has been established by solving the generalized dual-phase-lag (GDPL) equation \cite{Hooshmand20151}. The problem includes a non-homogeneous, time-dependent laser heat source. The GDPL is worked out by volume averaging the local instantaneous energy equation for blood and tissue { eliminating the blood temperature at the end. Then, the derived energy equation for the tissue temperature} can be treated by applying the separation of variables and Duhamel's superposition integral method for both absorbing and scattering tissues. The obtained temperature from the generalized DPL equation is lower than the one which is attained via the classical Pennes bioheat transfer. This reduction is attributed to contemplated heat convection between the blood vessels and the tissue. This research concludes that the generalized DPL equation reduces to the classical Pennes heat conduction equation by letting both $\tau_T$ and $\tau_q$ equal to zero. { This is straightforward as, in that case, the time delays disappear. Also, when both time delays are equal, Fourier's solution is recovered, similarly when Fourier resonance occurs in the GK equation.} This finding is in agreement with the work by Afrin \emph{et al.} \cite{Afrin2012} and is also emphasized in \cite{Ziaei2016}. 

Besides, two-dimensional exact analytical analysis of the Fourier/non-Fourier bioheat transfer equations for skin tissue subjected to an instantaneous heating condition is studied \cite{Askarizadeh2015}. The effects of blood perfusion and metabolic heat generation on thermal behavior are also investigated. A hybrid analytical scheme comprising the Laplace transform method in conjunction with the separation of variables technique and inversion theorem is developed, and the DPL and Pennes models are solved. It is established that the DPL bioheat transfer equation considering the effects of blood perfusion and metabolic heat generation reduces to the Pennes bioheat transfer equation when $\tau_q$ = $\tau_T$. The appearance of $\tau_T$ makes the heat diffusion in the tissue easier and consequently results in increased thermal damage. Simultaneously, burn time decreases while the thermal relaxation times on the BCs become negligible. In the GDPL model, based on the theory of porous media, the phase lag times are described in terms of the properties of the blood and tissue, interphase convective heat transfer quotient, and the blood perfusion rate.

The 1D bioheat transfer with the pulse boundary heat flux is investigated via the generalized dual-phase-lag model utilizing the hybrid application of the Laplace transform technique and the modified discretization technique \cite{Liu20151}. The obtained temperature distribution is far different from the one acquired by the classical DPL model and the Pennes equation. The authors affirmed that the GDPL equation is not reducible to the Pennes bioheat transfer one for $\tau_q=\tau_T$ or even for $\tau_q=\tau_T=0 $ s, in agreement with \cite{Liu20151}, but contrary to \cite{Afrin2012,Hooshmand20151}. This research \cite{Liu20151} claims that both phase lag times depend on various parameters of the problem, such as porosity, heat capacities of blood and tissues, coupling factors, and the ratio of thermal conductivity of the tissue and the blood are not independent. 

Furthermore, nonequilibrium effects during the laser irradiation of the living tissue have been studied in \cite{KCLiu20162}. Again, the hybrid Laplace transform and the modified discretization technique are employed to solve the problem. The dependence of the phase lag times on the porosity, heat capacities of blood and tissues, coupling factor, and the ratio of the thermal conductivity of the tissue and the blood are also considered.

Such reliance on the phase lag times has been missed in the previous studies in which the phase lag times are considered independently. Also, the transient bioheat transfer in 3D living tissue under the effect of the laser radiation internal heat source has been studied in \cite{MJasinski2016}. The generalized DPL model is used for the subdomain tissue, while the blood subdomain is modeled with an ordinary differential equation. The formulated GDPL problem is worked out employing the explicit scheme of the finite difference method. The calculation has been performed for different porosity values, the ratio of blood volume to the total volume. For a lower porosity, the temperatures and the degree of tissue damage found through the Arrhenius integral are more significant than that of the higher porosity. The blood perfusion rate in GDPL is presented in an implicit way through the convection-perfusion coupling factor. The thermally damaged (necrotic) tissue results in a zero perfusion rate, reducing the value of the coupling factor and the porosity value and consequently increasing the temperature. Moreover, the effective thermal conductivity is impacted by the $\tau_T$, $\tau_q$.

Unlike the local equilibrium-based models, the phase lag times are not independent for nonequilibrium situations. Hence, the generalized DPL model based on the nonequilibrium heat transfer model is developed for finding the temperature distribution during the electromagnetic radiation thermal therapy of the tumor inside the living tissue. The generalized DPL equation is solved using a hybrid numerical framework well-established on the finite difference scheme and the Chebyshev wavelet Galerkin method \cite{PKumar2019}. Additionally, it is mentioned that a small number of Chebyshev wavelet basis functions are sufficient for obtaining the desired accuracy. The authors state that the existing convection-perfusion coupling factor in the generalized DPL equation results in lower temperature at the position of the tumor in comparison to what is obtained from the hybrid model based on the classical DPL and the Pennes model. Also, more significant phase lag times cause lower temperatures at the tumor position.

It is further confirmed that the larger porosity and interfacial convective heat transfer yield lower temperatures. This recognition is absolutely in agreement with the one obtained in \cite{MJasinski2016}, where the lower porosity is found to result in higher temperature. Apart from that, the verification of the previous numerical findings ascertains the correctness of the present numerical procedure. Also, the nonequilibrium heat transfer in living biological tissues as porous mediums has been investigated analytically in the context of the generalized DPL model \cite{HAskarizadeh2015,Kaluza2017}.

%--------------------------------------------------------------------
\subsubsection{Generalized boundary-element method (GBEM)}
The heat transfer process in the 3D domain of bioheat tissues described via the DPL bioheat equation is modeled with the general boundary element method (GBEM) \cite{Majchrzak2015}. In GBEM, one of the types of the boundary element method, the derivatives of temperature with respect to time are substituted by the differential coefficients. The BEM itself decreases the problem's spatial dimensions by one and is also a method with high accuracy. Besides, the local/temporary boundary heat fluxes and temperature are directly calculated. The efficiency and exactness of the algorithm are verified by comparing the results for the special case with the one obtained from the analytical solution.

Recently, the GBEM has been developed for the DPL heat conduction model of 1D two-layered thin metal film \cite{Majchrzak2020}, and 3D bilayered microdomain \cite{EMajchrzak2019}. Furthermore, the bioheat heat transfer process described by the DPL equation and the appropriate boundary-initial value conditions has been investigated via the explicit scheme of the generalized finite difference method (GFDM) for the first time in \cite{LTurchan2017}. Most methods dealing with solving a partial differential equation are based on one of the classic variants of the finite difference method (FDM), the control volume method, the finite element method (FEM), or the boundary element method. The GFDM is, in a way, the bridge between the classical FDM and FEM, resulting in the possibility of arbitrary discretization of the domain. In \cite{LTurchan2017}, a cylindrical tissue with an internal heat source is considered. In the given domain, the cloud of nodes is created with an n-point star attributed to each node.
The n-point star is constructed by the central node and the surrounding points. The accuracy of numerical calculation is directly related to the point selection. Cases with various node density values and sizes of n-point stars have been numerically analyzed. Comparing the obtained results with the ones calculated from the classical finite difference method using fine mesh demonstrates a global error of less than one percent. 

%--------------------------------------------------------------------
\subsection{SPH method}
Smoothed-particle hydrodynamics is a particle-based numerical method, which has been extensively used for fluid mechanics or deforming domain problems. Ghazanfarian et al. \cite{Ghazanfarian20152} developed the SPH method to solve the DPL/CV-based bioheat model in the nonlinear regime. They found excellent agreement between the results of the SPH method and other numerical techniques. However, there is a potential to develop the SPH method for high-dimensional cases for bioheat non-Fourier problems, including coupled physical phenomena.

%--------------------------------------------------------------------
\subsubsection{Statistical method}
The thermal damage during the laser irradiation of the living biological tissues is studied in the framework of the GDPL. In more detail, a sample-based stochastic model has been used to deal with the GDPL problem. Although the remarkable advancement in modeling and simulation of laser therapy has been achieved, fewer studies are devoted to the unstable characteristics of the thermal effects directly caused by the intrinsic uncertainties of the input parameters and thermophysical properties \cite{Afrin2017}.

The uncertainties are investigated for the input parameters, including laser exposure time, blood perfusion, scattering coefficient, diffuse reflectance of light, and phase lag times, which all are considered to obey the Gaussian distributions of the uncertainty. When the distributions of input parameters are obtained, Monte Carlo Sampling (MCS) is used to find the combination of the input parameters. The input parameters, being randomly selected, are joined together as one sample.

In stochastic modeling, the {variabilities} of the output parameters is assessed in reliance on the uncertainty of the input parameters. Afrin and Zhang have obtained that the effects of laser exposure time, phase lag times, and the blood perfusion on the output parameters of maximum temperature and the thermal damage overshadow the efficacy of the scattering coefficient and the diffuse reflectance of light on the irradiated surface of tissues. Further, a surrogate-based optimization framework has been implied to optimize the thermal damage in the living biological tissues \cite{Afrin2019}. Every input variable individually quadratically responds to the output parameters for highly absorbing tissues.
{
%--------------------------------------------------------------------
\subsubsection{Higher-order models}
The thermal responses concerning nonequilibrium heat transport in biological tissue, taking into account the second-order effects in lagging, are studied in \cite{Liu20152}. Strictly speaking, the non-Fourier bioheat transfer equation containing the mixed-derivative terms and the higher-order derivatives of temperature relative to the time has been established.} To overcome the mathematical difficulties of this nonlinear DPL (NDPL) problem, a hybrid scheme of the Laplace transform, and the modified discretization technique \cite{KLiu2011} in conjunction with the hyperbolic shape functions is used to treat the present problem. It is confirmed that although the difference between the temperature obtained from the higher-order NDPL and the DPL equation seems negligible, there exists an apparent deviation within the computed results of the thermal damages via the Arrhenius equation. This is ascribed to the nonlinear behavior of the Arrhenius equation as a function of temperature.}

%--------------------------------------------------------------------
\subsubsection{Separation of variables techniques}
The temperature distribution and thermal damage in skin tissues under the pulse laser heating and fluid cooling are analyzed using the Pennes, MCV, and DPL models. Specifically, the bioheat DPL model considering the general boundary conditions is solved \cite{Lin2016}. The second-order differential equation with nonhomogeneous boundary conditions is dealt with by extending the shifting variable method firstly established in \cite{Lin1996}. In \cite{Lin2016}, the comparison of the thermal damage index, $\Omega$, obtained via several thermal damage models is performed. Most of the models have the form similar to the more famous Arrhenius burn integration proposed by Henriques and Moritz, $\Omega=\int_{0}^{t} \mathrm{A}\,\exp(-\mathrm{E}_a/\mathrm{R}T) \textrm{d}t$ with A, E$_a$, and R being, respectively, the material frequency factor, the activation energy, and the universal gas constant \cite{FXu2010ch}.

Hence, the difference between the models directly relates to the discrepancy of their coefficients which are implied in the burn damage integral. The difference between the coefficients is attributed to the different experimental databases used to delineate the models and also unlike focus while analyzing the burn process. For all three models (Pennes, MCV, and DPL) under the same condition, $\Omega$ of the Wu model is the largest while that of the Fugitt is the lowest. Also, the $\Omega$ of the Henriques model is overestimated due to the temperature-independent activation energy and the frequency factor. Further, the effect of the simultaneous implication of pulsed heating and the cooling of the skin for thermal therapy is studied. On the other hand, the exact analytical solution of temperature distribution in living tissues during the thermal treatment is studied by solving 1-D Pennes' bioheat equation (PBHE) with the separation of variables technique \cite{Dutta2017}.

The authors have claimed that most of the research has used the constant initial steady temperature, which is not entirely appropriate for the biological tissues for living cells. The authors have also asserted that the term metabolic heat generation in the PBHE has been ignored in all previous literature. Contrary to this claim, the calculations in \cite{Askarizadeh2015} include the metabolic heat generation term. This work includes metabolic heat generation and investigates the biological tissues analytically for two different spatially dependent and constant initial conditions. They compare the obtained temperature profile with the available experimental data, and it is affirmed that the spatially dependent steady-state initial condition is the best option for determining the thermal distribution in the living tissues. Here, we point out that nonhomogeneous initial conditions must be compatible with the constitutive equation and that compatibility determines the initial time derivative. This property is not investigated in general. Therefore, the related results must be treated with reservations.

%--------------------------------------------------------------------
\subsubsection{Fractional-order DPLBHT equation}
In 2015, the fractional dual-phase-lag model combined with the corresponding Pennes bioheat transfer equation was proposed \cite{XH-Y2015}. Fractional calculus is a concept defined for integrals and derivatives of arbitrary order. There exist various definitions of fractional order differentiation, such as Grunwald-Letnikov's \cite{Scherer2011}, Riemann-Liouville and Caputo's \cite{CLi2011} definition. Here, the Caputo fractional derivative is substituted in the fractional DPLBH equation. The fractional operators replace the two first-order time derivatives,
\begin{equation}
\nonumber
q(r,t)+\tau_{q}^{\alpha}\frac{\partial^{\alpha}}{\partial t^{\alpha}}q(r,t)=-k\{\nabla T(r,t)+\tau_{T}^{\gamma}\frac{\partial^{\gamma}}{\partial t^{\gamma}}\nabla T(r,t)\}
\end{equation}
and also the phase lags $\tau_q$ and $\tau_T$ are substituted by $\tau_{q}^{\alpha}$ and $\tau_{T}^{\gamma}$ to preserve the dimensions of the equation. The problem is analytically solved by applying the Laplace and Fourier transforms, and the solutions are presented through the Fox H-functions \cite{Mainardi2005}. By fitting the fractional DPL predictions to the available data for temperature distribution and using the nonlinear L–M least-square algorithm for experimental data fitting, two relaxation times and the orders of fractionality are found. Later, the thermal behavior in living biological tissues is investigated using the time-fractional DPLBHT model with a different technique \cite{Kumar20153,DKumar2017}. The living tissue is affected by the metabolic and electromagnetic heat sources during the thermal treatment and the Dirichlet boundary condition. 

To reduce the time-fractional DPLBHT equation into the system of ordinary differential equations, the spatial discretization framework in space is applied with initial conditions in vector-matrix form. Subsequently, the time-fractional ODEs are converted into the Sylvester matrix equation by using the finite element Legendre wavelet Galerkin method (FELWGM) with the block pulse function in terms of Caputo fractional order derivative. The present case's multi-resolution analysis of Legendre wavelets localizes small-scale variations of the solution and the fast switching of functional bases. Good consistency between the obtained results from the FELWGM and the exact solution is reported. In brief, the time-fractional order derivative as an important parameter in thermal treatment of the cancerous tissues to accurately handle the temperature is the objective of the mentioned paper. The 2D transient, DPL, and variable-order fractional energy equation form of the bioheat transfer equation are studied in \cite{Hosseininia2019}. The semi-discrete method based on the two-dimensional Legendre wavelets (2D LWs) has been used. In brief, the main problem is reduced to the easily solvable system of algebraic equations. Three different $\alpha$ are investigated from 0.5 to 1.0 and from 1.0 to 0.5. It is reported that increasing $\alpha$ brings up more intense temperature peaks earlier. On the other hand, the constant time-fractional order leads to a more uniform temperature distribution. Presenting no verification with any of the available data, the authors claim that their method's stability and spectral accuracy are checked with their results obtained for a numerical experimental example. Increasing the fractional-order from $\alpha=$0.1 to $\alpha=$1.0 causes the augmentation of the maximum temperature of the tissue for about 29$\%$. Concerning this paper, serious and critical problems are signified in a recently published comment \cite{Pantokratoras2020}. It is pointed out that the main equation in \cite{Hosseininia2019},
%\begin{widetext}
\begin{equation}
\rho c \frac{\partial^{\alpha}T(x,t)}{\partial t^{\alpha}}=k\frac{\partial^{\beta} T(x,t)}{\partial x^{\beta}}+\mathrm{Q}_{metabolic}+\mathrm{Q}_{perfusion}+\mathrm{Q}_{source},
\end{equation}
%\end{widetext}
is dimensionally problematic. In the mentioned paper, it is stated that the above equation holds for $0< \alpha \leq 1$ and $0< \beta \leq 2$. However, to put it more precisely, it is obvious that only the two first terms in the equation are dimensionally matched for the case with $\alpha$ = 1 and $\beta$= 2. Even for this special case, the above equation is still wrong as the unit of the term [Q$_{perfusion}$]= kg m$^2$sec$^{-3}$ differs from the others such as [Q$_{metabolic}$]=kg m$^{-1}$sec$^{-3}$. Further, the two defined dimensionless parameters, P$_f$ and P$_m$, are shown to be evidently dimensional. The units of $\mathrm{P}_F=\sqrt{\frac{\omega_b c_b \rho_b}{k}}$ and $\mathrm{P}_{m}=\frac{Q_m L^2}{k}$ are respectively, m$^{1/2}$ and K. These points cast doubt on the correctness of the results that appeared in \cite{Hosseininia2019}.
\newline
A space and time-fractional DPL bioheat transfer model in the presence of temperature-dependent metabolic and space-time dependent electromagnetic heat sources has been developed \cite{KNRai2020}. The fractional-order partial differential equation is again reduced into the system of algebraic equations by implying the Legendre wavelet collocation method. The set of equations is solved using the Newton iteration method. The error bound and stability analysis and the numerical method validation by comparison with the exact solution are presented. To annihilate the cancerous tissues while keeping the healthy cells around intact, finding the accurate temperature distribution in the tumor zone is a necessity.

Accordingly, the effect of variability of time/space fractional derivative order, the transferred power, and the phase-lagging times on the temperature distribution are investigated. The best choices, providing the desired temperature at a particular time and space, are found. In more detail, the tissue temperature increases as the space-fractional order derivative grows, and decreases when the value of the time-fractional order derivative reduces. Further, the effect of the $\tau_q$ is reported to be more noticeable than the $\tau_T$. It should be noted that the comment \cite{Pantokratoras2020} questioning the non-dimensionality of the parameters P$_f$ and P$_m$ also apply for the papers \cite{DKumar2017,KNRai2020}.

At this time, it is noteworthy to mention a new hybrid algorithm based on integrating the local radial basis function collocation method (LRBFCM) and the general boundary element method for dealing with time-fractional DPLBHT problems in functionally graded tissues proposed in 2019 \cite{MAFahmy2019}. The new theory claims that the general solution (T) of the time-fractional DPLBHT equation is the sum of the T$_f$, the solution of the fractional-order governing equation without phase lags, and T$_d$, the solution of the DPL governing equation without fractional derivative, say T=T$_f$+T$_d$. Caputo's time-fractional derivative is used to replace the fractional PBHT equation with a series of integer-order PDEs. Correspondingly, the LRBFCM is implied to deal with the obtained PDEs at discrete time steps. On the other hand, the DPL equation without considering the fractional-order derivative is resolved via GBEM. It is worth noting that the envisaged theory can be utilized to treat the time fractional-order BHT problems with single or triple-phase-lag as well.

The validity and accuracy of the suggested procedure are verified via comparison of the LRBFCM-GBEM obtained results with that of the finite difference method \cite{LTurchan2017} and finite element method \cite{DKumar2017}. A new constitutive model, based on the combination and development of the Cattaneo-Christov model and DPL model, is suggested to investigate the macroscopic and microscopic heat transfer in the moving media \cite{LLiu2018}. The governing equation includes the phase lag times and the time-fractional derivative with the highest order of 1+$\alpha$ (0<$\alpha \le$1). Solutions are calculated numerically utilizing the L1 scheme, which is one of the most reputable and prosperous numerical methods for discretizing the Caputo fractional derivative in time. The model is used to evaluate the heat conduction in processed meat and find out the effects of the convection velocity on the temporal evolution of the temperature. It is concluded that a larger positive convection velocity leads to slower temperature transport. Also, the thermal behavior of biological tissue has been investigated.

It is obtained that the faster temperature transport occurs at smaller positions while it gets slower for larger positions when the fractional parameters, $\alpha$, and $\beta$ are respectively larger and smaller. Also, the temperature transport becomes slower for a larger macroscopic phase lag time or the convection velocity. Furthermore, the super-diffusion fractional Cattaneo heat conduction model has been applied to simulate the heat diffusion through the skin tissue with a heat source \cite{Goudarzi2019}. In further detail, the three-layered skin tissue, in contact with a hot water source and a single-layer skin tissue exposed abruptly to a laser heat source, is investigated. The governing equation is solved using an implicit method and a finite volume technique for the first and second cases. The results of the Fractional Single-Phase Lag (FSPL) model have been compared with the available data of DPL modeling. More concretely, the temperature profile for the FSPL model with $\alpha$=0.9985 and the phase lag, $\tau$=16 s, affected by laser on the boundary agrees well with the result obtained from DPL with phase lag times of $\tau_q=16$ s and $\tau_T=0.05$ s.

%--------------------------------------------------------------------
\subsubsection{Inverse problems}
The inverse hyperbolic and DPL heat conduction problems for estimation of the unknown time-dependent laser irradiance and the thermal damage in laser-irradiated three-layer skin tissue are respectively investigated in \cite{HLLee2015} and \cite{YCYang2017}. The inverse algorithm is based on the conjugate gradient method (CGM)and the discrepancy principle. CGM is deducted from the perturbation principles and converts the inverse problem to the solution of three other problems called as direct, sensitivity, and adjoint. The obtained results reveal that the evolved method accurately predicts the unidentified laser irradiation.

It is declared that the developed technique performs the inverse calculation needless of any initial guess for the functional form of the indeterminate quantities. While the laser irradiation is assessed, the particular temperature profile and the thermal damage can be attained at the irradiated surface. Also, the solution of the inverse problem in predicting the laser intensity in biological tissues via gradient method is presented in \cite{Majchrzak2019}. As the gradient algorithm is not always convergent, selecting an appropriate starting point is crucial to ensure convergence. Besides, the phase lags $\tau_q$ and $\tau_T$, and the thermal diffusivity $\alpha$ of processed meat are calculated based on the DPL mode \cite{KCLiu2016}. The Laplace transform and the least-squares scheme are used to estimate the unknown parameters. The discrepancy between the predicted temperature and the experimental data is reduced using the least-squares minimization method.

At last, it is also interesting to mention the work by Ismailov \emph{et al.} where the time-dependent blood perfusion coefficient for the conventional Pennes bioheat equation with Ionkin-type nonlocal boundary and integral energy over-determination conditions has been estimated \cite{Ismailov2018}. In the previous techniques, the original problem is converted into the inverse source problem, and consequently, the perfusion coefficient is predicted within the numerical differentiation. In contrast, the coefficient is predicted directly via a nonlinear minimization technique. In more detail, the method of lines based on a highly accurate pseudospectral approach has been used to resolve the bioheat equation, and the perfusion coefficients are found by the Levenberg–Marquardt method using the discrepancy principle as a stopping rule.

%--------------------------------------------------------------------
\subsubsection{Transient radiative heat transfer}
In this section, the studies modeling the light-tissue interaction using the DPL coupled with the radiative transfer equation (RTE) in different frameworks are presented.
\paragraph{Discrete ordinate method (DOM)}
The DPL based heat conduction model coupled with the transient form of RTE for investigating the phenomenon of light propagation inside the tissue phantom is developed in \cite{SKumar2015}. Two-dimensional distribution of the light intensity inside the tissue is found by solving the RTE using the discrete ordinate method (DOM). The heat transfer is modeled using the finite volume method (FVM) based discretization.

After that, the temperature distribution inside the biological tissue phantom embedded with optical inhomogeneities of varying contrast levels has been determined using the DPL-based model. This study presents complete and comprehensive verification of the numerical code and compares the oscillation's temperature distribution and magnitude for three different heat conduction models named hyperbolic, DPL, and Fourier. The DPL temperature profile lies between that of the hyperbolic and Fourier models. Moreover, the authors confirmed that the higher magnitude of the oscillations is attributed to the hyperbolic modeling due to the effect of the $\tau_q$. On the other hand, the DPL model predicts smaller oscillations owing to the coupled effects of $\tau_q$ and $\tau_T$.
\paragraph{Lattice Boltzmann method (LBM)-based numerical framework}
Further, the thermal response of the biological tissue under the effect of the laser irradiation, the coupled transient RTE, and DPL heat conduction equation has been firstly solved via developing and utilizing a Lattice Boltzmann method (LBM)-based numerical framework in \cite{SPatidar2016}. The intensity distribution inside the tissue phantom has been calculated by working out the transient RTE in more detail.

In other words, the integro-differential equation is transformed into a set of PDE conforming to the finite number of lattice directions M by using the D2Q8 model in LBM. Then the solution is coupled with a generalized DPL model. The uniform solver based on the concept of LBM has been found to anticipate the thermal responses of the laser-irradiated biological tissue adequately. Additionally, the heat transfer through the biological tissue phantoms has been numerically modeled in a cylindrical coordinate system \cite{Sravan2018}.

\paragraph{Finite integral transform scheme}
Kumar and Srivastava argued that the complete FIT-based analytical solution of the DPL bioheat equation, being subsequently coupled with the numerical results of transient RTE, has been developed \cite{Kumar20171}. The temperature distribution is calculated for three different time-independent/dependent (sinusoidal) BCs and short pulse laser irradiation.

%--------------------------------------------------------------------
\subsubsection{Green function technique for porous media}
The rapid heating of living biological tissues during the hyperthermia treatment is explained with the assessment of the two-equation (medium or step) model \cite{Monte20172,Monte2017}. The two-step model considers the biological tissue like a porous medium under Local Thermal Non-Equilibrium (LTNE) circumstances \cite{Minkowycz1999}. This model is based on the two coupled partial differential equations written for the tissue (solid phase) and the blood (fluid phase). An adequate transformation of the dependent variable leads to the cancellation of the bioheat term and then the transient DPL diffusion equation will be analytically solved employing the DPL-based alternative Green’s functions solution equation (AGFSE).

As the solid phase temperature is known, the blood temperature is calculated utilizing the approximate lumped capacitance analysis, which reduces a thermal system to a number of discrete lumps with no temperature difference inside for porous media and the particular application of the biological tissues. The blood temperature is found to be retarded due to the relaxation time between the two solid and fluid phases. Further, larger values of the porosity and perfusion rate of the tissue improve the cooling influences of the blood through the vessels and lower the thermal damages. Additionally, the exact series solutions for finding the temperature distribution in porous passages are developed in \cite{AHji-Sheikh2020}. Special attention has been paid to the rapid heating through the biological systems, and also the parallel plate and the circular porous passages, filled with solids, have been investigated.

%--------------------------------------------------------------------
\subsubsection{Different heating sources}
This section discusses the recent literature presenting the effect of the various kinds of heating sources. Remarkably, the time and space-dependent heat sources and moving heating sources treated with different methods are explored.
\paragraph{Time/space dependent heating source}
At first, the results for the media under the effect of the time and space-dependent heat flux are presented \cite{KNRai2020}. Recent works have treated this problem utilizing the following methods.
\subparagraph{Local thermal non-equilibrium (LTNE) approach}
The two-dimensional local thermal non-equilibrium bioheat model (LTNE-DPL) has been developed to investigate two-dimensional malignant tissues under a hyperthermia treatment \cite{Dutta2019}. As the skin tissues are porous structures composed of a highly non-uniform non-homogeneous fluid and solid medium structure, an LTNE model is preferred over a local thermal equilibrium process. The LTNE-DPL has been solved analytically by applying a hybrid scheme based on the change of the variables and FIT with heat flux boundary conditions and spatial-dependent initial conditions.

Two biological cases with imposed oscillating and constant heat flux on the diseased tissue have been investigated. The sinusoidal therapeutic heat flux is found to be better due to its longer time of therapy relative to the constant heat flux heating. This long exposure time results in the destruction of more cancer cells. Also, the authors employ a medium range of tissue heating of 38–44$^\circ$C for the lengthy thermal therapy to avoid the occurrence of the internal injury, instead of applying the usual average temperature of 50$^\circ$C for 30 mins.

\subparagraph{Radial basis function (RBF) approximations}
The heat distribution in skin tissue under the influence of the constant and sinusoidal heat flux at the surface has been studied in \cite{Verma2020}. The DPL bioheat transfer model is solved using the finite difference and radial basis function (RBF) approximations. It is found that decreasing the heating frequency of sinusoidal heating leads to a higher amplitude of the thermal wave. Also, healthy tissue damage is less probable during the sinusoidal heat flux condition than that of the constant heat flux condition. This is in agreement with what is obtained in \cite{Dutta2019}.

\paragraph{Moving heat sources}
Secondly, the cases with implied moving heat sources are considered. Recent studies treat such affected mediums with Green's function technique and FIT.
\subparagraph{Green's function technique}
A 3D cuboid biological tissue, subjected to a time and space-dependent moving laser heat source, has been modeled in \cite{Ma20182}. The Green function for the three-dimensional DPL bioheat equation has been derived, and it is used to find the temperature distribution. It has been found that the peak temperature emerges at the zone being directly irradiated by the laser beam, and its position moves with the laser beam. Furthermore, the peak temperature value is inversely dependent on the speed of the laser beam. The smaller spot size of laser beam results in much concentrated laser power, which causes extreme temperature growth in a small affected area.

\subparagraph{Finite integral transform method}
An exact analytical investigation of the heat transfer in a 3D square-shaped cuboid plate under the effect of a moving laser heat source has been performed based on the combined form of FIT and Duhamel's theorem \cite{JDutta2020}. A time-decaying laser heat source term has been considered. Duhamel's theorem is employed due to the existence of the time and space-dependent heat source term in the mentioned DPL heat conduction model. This theorem specifically transforms the non-homogeneous problem with time-dependent heat source and constant initial conditions to a homogeneous one with time-dependent initial conditions. Further, the temperature response is found to be highly dependent on the laser power density and duration of the laser exposure. In more detail, the increase in laser power density results in a larger peak temperature. Also, the higher laser exposure time makes the maximum temperature decrease. The finding is in contrast to what was obtained for the generalized DPL bioheat equation in \cite{Afrin2017}. This conflict can be attributed to the non-biological, biological nature of the problems and the difference between investigated systems' dimensions. The present study considers 3D plates while 1D living biological cases are studied in \cite{Afrin2017}.

%--------------------------------------------------------------------
\subsubsection{Eigenfunction-based solutions for time-dependent BC}
It is also interesting to bring up the new research studying the cases with time-dependent boundary conditions. Recently, the DPL bioheat transfer problem with time-dependent BCs for a 2D skin tissue phantom is solved analytically based on the finite integral transform \cite{Abdel-Hamid1999}. Biswas \emph{et al.} have developed a technique of homogenization of boundary conditions to eradicate a possible mismatch between the BCs and corresponding eigenfunctions \cite{Biswas20201}. The orthogonal eigenfunction expansion method (OEEM) is employed with the new homogenization technique to the DPLBHT model. The homogenization of the generalized time-dependent BCs is performed by subtracting an auxiliary function from the temperature of the domain of interest. Consequently, a problem in terms of the modified temperature with a modified source and homogeneous BCs is obtained. It is mentioned that an extra condition, say the pseudo-steady-state, is enforced to acquire quite simple, unique auxiliary functions. The cases with constant surface temperature and sinusoidal heat flux on the surface are studied. The aforementioned procedure takes away the large spurious oscillations nearby the boundary for constant surface temperature. In the case of the surface heat flux condition, the temperature obtained applying the homogenization approach is found to be in good agreement with that of the FIT approach. Also, the realistic non-zero flux distribution is obtained on the boundary, while the FIT procedure predicts zero heat flux when the same non-zero heat flux at the boundary is applied.

%--------------------------------------------------------------------
\subsubsection{Numerical toolboxes}
In the end, we concentrate on the newly developed user-friendly mathematical tools for solving the non-Fourier DPL equation. This toolbox is becoming very popular in any aspect of the numerical sciences, it is easy to use, needs less experience, and is mostly faster than the old-fashioned standard coding. Here we review the recent works involving the development of such tools.

\paragraph{Adaptive time integrators}
A new, low-cost, and accurate time marching technique is developed to deal with the DPL equation in \cite{Soares2018}. The spatial discretization of the model is performed using the FEM, and the temporal discretization is treated via a new algorithm. Hence, according to the characteristics and the calculated results, the adaptive time integration parameters locally adjust themselves spatially and temporally during the solution procedure. It has been demonstrated that this method is formulated as a non-iterative single-step process, and consequently, it is computationally low-cost. 

\paragraph{OpenFOAM solver}
One of the complexities of biological cases is their sophisticated geometrical details. Thus, to produce realistic results, we need to perform simulations using object-oriented open-source solvers. Regardless of the rapid growth of the non-Fourier heat transfer models, some numerical restrictions for real cases are still problematic. Jamshidi and Ghazanfarian \cite{MJamshidi2017,MJamshidi20182} developed an OpenFOAM solver to include additional terms of the DPL model with numerical flexibilities such as supporting structured and unstructured meshes and easy implementation of three-dimensional cases. 

Jamshidi and Ghazanfarian \cite{MJamshidi2020} simulated a more realistic three-dimensional multilayered skin of a human finger with buried embedded vessels based on the DPL model in the tissue. They developed a novel solver over the platform of OpenFOAM codes to model the effect of capillaries, small arteries, small veins, the blood velocity, the blood pressure, and the distance from the skin surface in the epidermis, the dermis, and the hypodermis layers. The temperature distribution and the peak temperature rise in the skin layers and the buried vessels are computed. Such solvers can be helpful to conduct ab initio simulations.

%-----------------------------------------------------------------------------------
\section{The roadmap for future works}
As we have seen, there is a gap between theory and practice in bioheat non-Fourier research. On the one hand, the practice requires knowledge of the possible heat source terms and perfusion properties of various tissues. Also, nonequilibrium thermodynamics and the dual-phase-lag approach provide modifications for the constitutive part. These are different parts of the continuum model, and for better understanding, the experimental separation of various physical phenomena is strongly required. Otherwise, one could not decide why the Fourier law does not explain the experimental data: is it lacking some important source terms, such as included in the Pennes' model, or is the Fourier law itself inadequate?
The theoretical developments lead to better insight with more reliable, stable, and solvable models. The practical requirements may be less consistent but better suited to the immediate need of the particular problems. The careful experimentation with comparative modeling can lead to the bridge over the gap between the various theories. The practical side needs more accurate models with easy implementation. This is contradictory with developing such a specific model in which one needs to know the velocity field of the blood flow, the exact structure of the artery-vein pairs, or other particular properties of the tissues.

From a theoretical point of view, we have seen that the dual-phase-lag concept is an easy way to get partial differential equations of non-Fourier heat conduction. However, the system is too general, and boundary conditions are necessary for solutions and experimentation. Moreover, one can encounter instabilities due to the missing thermodynamic background. On top of this, the nonlinearities by temperature-dependent parameters are important for practice, and that property cannot be consistently included in a DPL approach. There is a need for models with stronger physical background, providing modeling capabilities as general as possible without losing accuracy and reliability. Such models with strong background are the Jeffreys and Guyer-Krumhansl equation. Although they similar to the DPL model, and in the linear regime, they all have the same temperature representation, they differ in the constitutive part, and must not be mixed with the DPL equation. The necessary background is given by nonequilibrium thermodynamics, as it was demonstrated in Sections 2 and 3. It is remarkable that the experimentally fitted relaxation times of DPL modeling all fulfill the theoretically expected stability conditions in bioheat transfer.

The research in non-Fourier bioheat transfer can be assorted in three subcategories. The first sub-class includes studies that are involved with modeling the therapy to suggest the conditions leading to the optimum thermal treatment with maximum necrosis of the malfunctioning tissue and minimum change of the healthy ones. Despite numerous research on this topic, there are still two open paths for future studies. Although bioheat transport has been studied for many biological cases, such as modeling the benign and malignant cancerous tumors in organs like a tooth, thyroid nodule, skin, breast, and liver, there are hardly any studies dealing with the thermal remedy of abnormality in the brain or heart organs.

To be more precise, ablation treatments have been utilized for neurological brain disorders to create curative lesions through unfunctional brain circuits, annihilate intracranial tumors, and space-occupying masses. The ablative methods used for brain surgery are RF thermoablation, stereotactic radiosurgery (SRS), laser interstitial thermal treatment, and magnetic resonance-guided focused ultrasound (MRgFUS). In addition, as mentioned before, RF ablation and cryoablation are the usual techniques for treating heart arrhythmia. Cryoablation surgery being more safe and effective for the children and young adults patients \cite{Hanninen2013} has not been yet modeled. As these operations are so case-sensitive and should be performed very precisely, the prediction of temperature demeanor to get aware of the patient's condition during the surgery is vital. Hence, one future prospect is to extract the studies to include the non-Fourier modeling of the heat conduction during these mentioned surgeries and similar cases.

Another challenge for suggesting the appropriate situation to facilitate the result of the therapy technique is to bring the situation as close as possible to the real conditions. This can be done by turning the modeling cases into more realistic three-dimensional ones or by providing the calculations' temperature, position, or time-dependent parameters. 
Accordingly, the future works will be mostly based on three-dimensional modelings in which the involved known parameters are not constant. Also, considering unstable thermal effects directly induced by the intrinsic uncertainties of the input characteristics and thermophysical, which results in more precise modeling, is a step forward.

The second subcategory is involved with the pain management of the patients. The skin's thermo-mechanical responses throughout the therapy are directly related to the distress that the patient endures.
More particularly, the integrated analytical context, including the thermo-mechanical model; the scheme for nociceptor transduction like the Hodgkin-Huxley form model; and pain modulation and perception modeling such as gate control theory; have to be developed to quantify the skin pain sensation. As pain management ensued from the obtained predictions, it leads to the less irritating clinical treatments; the studies in this topic will be of great importance. In other words, being sure of the patients' safety, the research in subsequent stages should be dedicated to providing the patient comfort.

The following important sub-class comprises the burn issues and also the protective clothing. Despite the many research results dealing with skin burn, only a few studies have been devoted to appropriately investigating protective clothing. The protective fabrics between the heat source and skin make the usual methods to inaccurate predictions as the incident heat can be modified in value and form crossing the garment. Scrutinies in this subject are directly related to the skin bio-thermomechanics, which are advantageous for medical usage and are also noteworthy in space and military missions. The harsh and intense circumstances experienced in space travel and military operations make the procurement of desirable clothes for thermal protection of the astronauts and army personnel; an inevitable problem. Hence, more research on the topic of bio-thermomechanics should be performed.

The number of studies on nanomaterials and nanoparticles that are great candidates for different biomedical therapies, including activated hyperthermia, is not too much. More research on this topic will lead the researchers to suggest more suitable nanomaterials and help the clinical physicians control the situation better during the therapy. For instance, one of the potential nanomaterials known as carbon nanotube (CNT), a nested, cylindrical graphene structure and a diameter ranging from a few to hundreds of nanometers, can be an exciting topic of investigation.

On the other hand, cryopreservation, a procedure of preserving the biological tissues with the least severe failures by exposing the tissues to the very low cryogenic temperature, has gathered interest. The more obvious query improves the model to present the more realistic three-dimensional cases. Another challenge in the topic of non-Fourier DPL heat conduction modeling is the lack of information on the values of the $\tau_q$ and $\tau_T$. Although the DPL equation was proposed in 1995, no accurate prediction of the phase lag times in bioheat tissues has been confirmed.

Thereby, the focus should also be on finding the real values of the delay times. The inaccurate phase lag times are also dependent on the porosity, heat capacity of the blood, and the coupling factor. Such reliance on phase lag times, which also affects the tissue's thermal response, is missed out in most of the previous publications and needs to be involved for getting more precious results. At last, it should be noted that only a few works present strong verification. That is why theoretical groups must contact the experimentalists and the clinical physicians to provide the available data and experience to certify the calculations.

\bmhead{Acknowledgments}

The second and the third authors thank ROCKSTUDY Ltd. (K\H om\'er\H o Kft.), Hungary led by L\'aszl\'o Kov\'acs for producing the rock samples. The research reported in this paper and carried out at BME has been supported by the grants National Research, Development and Innovation Office-NKFIH FK 134277, K 124366, and by the NRDI Fund (TKP2020 NC, Grant No. BME-NCS) based on the charter of bolster issued by the NRDI Office under the auspices of the Ministry for Innovation and Technology. This paper was supported by the J\'anos Bolyai Research Scholarship of the Hungarian Academy of Sciences (R. K.). Supported by the ÚNKP-21-5-BME-368 New National Excellence Program of the Ministry for Innovation and Technology from the source of the National Research, Development and Innovation Fund (R.~K.). The first and the fourth authors thank MES of Russia. This work was carried out with the financial support of the Ministry of Science and Higher Education of the Russian Federation within the framework of the base part of the state assignment (no. 0778-2020-0005) (Z. S and I. V. K.).

%\section*{Declarations}

%Some journals require declarations to be submitted in a standardized format. Please check the Instructions for Authors of the journal to which you are submitting to see if you need to complete this section. If yes, your manuscript must contain the following sections under the heading `Declarations':


\end{document}